\documentclass[aps,prb,a4paper,10pt,twocolumn,showpacs,floatfix,superscriptaddress,preprintnumbers,longbibliography]{revtex4-2}
\usepackage{float}
\usepackage{dcolumn,graphicx,color,booktabs,microtype,afterpage}
\usepackage{amsmath}
\usepackage[charter,greekuppercase=italicized]{mathdesign}
\usepackage[mathlines]{lineno}

\usepackage{nicefrac}

\usepackage[overload]{textcase}

\graphicspath{{./}{figure/}}
\renewcommand{\tablename}{Table}
\makeatletter\renewcommand{\fnum@figure}[1]{\figurename~\thefigure.~}\makeatother
\makeatletter\renewcommand{\fnum@table}[1]{\tablename~\thetable.}\makeatother

\newcount\hh \newcount\mm
\hh=\time \divide\hh by 60
\mm=\hh \multiply\mm by 60 \mm=-\mm
\advance\mm by \time
\def\now{\number\hh:\ifnum\mm<10{}0\fi\number\mm}

\usepackage[colorlinks,plainpages=false,linkcolor=blue,urlcolor=blue,citecolor=blue,pdfpagemode=UseNone,pdfstartview=FitBH]{hyperref}

\newcommand{\ZR}{Zr$_5$Pt$_3$}

\newcommand{\tcr}[1]{\textcolor{black}{#1}}

\begin{document}

\makeatletter\renewcommand{\ps@plain}{%
\def\@evenhead{\hfill\itshape\rightmark}%
\def\@oddhead{\itshape\leftmark\hfill}%
\renewcommand{\@evenfoot}{\hfill\small{--~\thepage~--}\hfill}%
\renewcommand{\@oddfoot}{\hfill\small{--~\thepage~--}\hfill}%
}\makeatother\pagestyle{plain}

%\preprint{\textit{Preprint: \today, \now.}} %For internal use only, do not distribute.}}
%\linenumbers

\title{Fully-gapped superconducting state in interstitial-carbon-doped 
\texorpdfstring{Zr$_5$Pt$_3$}{Zr5Pt3}}
%Zr$_5$Pt$_3$C$_x$
%\title{\texorpdfstring{${}^{195}$P\MakeLowercase{t}}{195Pt} NMR
%investigation of \texorpdfstring{Zr$_5$Pt$_3$}{Zr5Pt3} and
%\texorpdfstring{Zr$_5$Pt$_3$C$_{0.3}$}{Zr5Pt3C0.3}}%\ZR\ and \ZRC}
%

%
\author{T.\ Shang}\thanks{These authors contributed equally}\email[Corresponding author:\\]{tshang@phy.ecnu.edu.cn}
\affiliation{Key Laboratory of Polar Materials and Devices (MOE), School of Physics and Electronic Science, East China Normal University, Shanghai 200241, China}

\author{J.\ Philippe}\thanks{These authors contributed equally}
\affiliation{Laboratorium f\"ur Festk\"orperphysik, ETH Z\"urich, CH-8093 Zurich, Switzerland}

\author{X.\ Y.\ Zhu}
\affiliation{Key Laboratory of Polar Materials and Devices (MOE), School of Physics and Electronic Science, East China Normal University, Shanghai 200241, China}
\author{H.\ Zhang}
\affiliation{Key Laboratory of Polar Materials and Devices (MOE), School of Physics and Electronic Science, East China Normal University, Shanghai 200241, China}
\author{B.\ C.\ Yu}
\affiliation{Key Laboratory of Polar Materials and Devices (MOE), School of Physics and Electronic Science, East China Normal University, Shanghai 200241, China}
\author{Z.\ X.\ Zhen}
\affiliation{Key Laboratory of Polar Materials and Devices (MOE), School of Physics and Electronic Science, East China Normal University, Shanghai 200241, China}
%
%
%\author{Y.\ Xu}
%\affiliation{Key Laboratory of Polar Materials and Devices (MOE), School of Physics and Electronic Science, East China Normal University, Shanghai 200241, China}
%
%
%\author{Q.\ F.\ Zhan}
%\affiliation{Key Laboratory of Polar Materials and Devices (MOE), School of Physics and Electronic Science, East China Normal University, Shanghai 200241, China}
%

\author{H.-R.\ Ott}
\affiliation{Laboratorium f\"ur Festk\"orperphysik, ETH Z\"urich, CH-8093 Zurich, Switzerland}

\author{J.\ Kitagawa} 
\affiliation{Department of Electrical Engineering, Fukuoka Institute of Technology, Fukuoka 811-0295, Japan}

\author{T.\ Shiroka}\email[Corresponding author:\\]{tshiroka@phys.ethz.ch}
\affiliation{Laboratorium f\"ur Festk\"orperphysik, ETH Z\"urich, CH-8093 Zurich, Switzerland}
\affiliation{Laboratory for Muon-Spin Spectroscopy, Paul Scherrer Institut, CH-5232 Villigen PSI, Switzerland}

\begin{abstract}
We report a comprehensive study of the Zr$_5$Pt$_3$C$_x$ superconductors, 
with interstitial carbon comprised between 0 and 0.3.
At a macroscopic level, their superconductivity, with $T_c$ 
ranging from 4.5 to 6.3\,K,  
was investigated via electrical-resistivity-, magnetic-susceptibility-,
and specific-heat measurements. 
The upper critical fields $\mu_0H_\mathrm{c2}$ $\sim$ 7\,T were determined
mostly from measurements of the electrical resistivity in 
applied magnetic fields. 
The microscopic electronic properties were investigated by means of 
muon-spin rotation and relaxation ($\mu$SR) and nuclear magnetic 
resonance (NMR) techniques. 
In the normal state, NMR relaxation data indicate an almost
ideal metallic behavior, confirmed by band-structure calculations, 
which suggest a relatively high electronic density of states
at the Fermi level, dominated by the Zr 4$d$ orbitals.
The low-temperature superfluid density, obtained via transverse-field $\mu$SR, suggests a fully-gapped superconducting state in Zr$_5$Pt$_3$ and Zr$_5$Pt$_3$C$_{0.3}$,
with a zero-temperature gap $\Delta_0$ = 1.20 and 0.60\,meV and a 
magnetic penetration depth $\lambda_0$ = 333 and 493\,nm, respectively. 
The exponential dependence of the NMR relaxation rates below $T_c$ 
further supports a nodeless superconductivity.
The absence of spontaneous magnetic fields below the onset of superconductivity, as determined from zero-field $\mu$SR measurements, confirms a preserved
time-reversal symmetry in the superconducting state of Zr$_5$Pt$_3$C$_x$.
In contrast to a previous study, our
$\mu$SR and NMR results suggest a conventional superconductivity 
in the Zr$_5$Pt$_3$C$_x$ family, independent of the C content.   
\end{abstract}

\maketitle\enlargethispage{3pt}

\vspace{-5pt}

\section{Introduction}\enlargethispage{8pt}
\label{sec:Introduction}
The $T_5$$M_3$ family, where $T$ is a $d$-transition or rare-earth metal and $M$ a (post)-transition metal or a metalloid
element, features three distinct structural symmetries: orthorhombic Yb$_5$Sb$_3$-type ($Pnma$, No.~62),
tetragonal Cr$_5$B$_3$-type ($I4/mcm$, No.~140), and hexagonal Mn$_5$Si$_3$-type ($P6_3/mcm$, No.~193). 
The Cr$_5$B$_3$-type structure is adopted by a broad range of binary and ternary compounds, e.g., the layered ternary compounds of transition metals with boron and silicon, with a $T_5$$X$B$_2$ stoichiometry ($X$ = P or Si),  
known to exhibit many interesting properties.
For $T$ = 3$d$-Mn or Fe, both $T_5$SiB$_2$ and $T_5$PB$_2$ are ferromagnets 
with Curie temperatures above 480 and 600\,K, respectively. Hence, currently they are being considered for room-temperature magnetocaloric applications or as rare-earth-free permanent magnets~\cite{Almeida2009,Xie2010,McGuire2015,Lamichhane2016}. 
For $T$ = 4$d$-Nb, Mo, or 5$d$-Ta, W metals, all $T_5$SiB$_2$ 
are superconductors, with transition temperatures in the 5 to 8\,K range~\cite{Brauner2009,Machado2011,Fukuma2011,Fukuma2012}. 
Furthermore, the recently synthesized tetragonal Mo$_5$PB$_2$ was shown to exhibit multigap superconductivity (SC) with a critical temperature $T_c = 9.2$\,K~\cite{McGuire2016,Shang2020}, the highest $T_c$ recorded in 
a Cr$_5$B$_3$-type compound. 

The hexagonal Mn$_5$Si$_3$-type structure possesses an interstitial 
2$b$ site, well suited for the intercalation of light elements,  
such as oxygen, boron, or carbon, employed to engineer the band 
topology and, ultimately, the electronic
properties. 
Superconductivity has been reported in several families of materials, 
including Nb$_5$Ir$_3$O$_x$~\cite{Zhang2017}, (Nb,Zr)$_5$Pt$_3$O~\cite{Cort1982,Hamamoto2018}, Nb$_5$Ge$_3$C$_x$~\cite{Claeson1977,Bortolozo2012}, 
or Zr$_5$Pt$_3$C$_x$~\cite{Renosto2019}, with the highest superconducting transition temperature $T_c$ reaching $\sim$ 15\,K.
Upon intercalation of oxygen~\cite{Zhang2017}, the $T_c$ of 
Nb$_5$Ir$_3$ increases up to 10.5\,K, while upon Pt doping, a 
crossover from multiple- to single-gap SC occurs in 
Nb$_5$Ir$_{3-x}$Pt$_x$O~\cite{Kitagawa2020,Xu2020}. 
Contrary to the Nb$_5$Ir$_3$O$_x$ case, in Zr$_5$Pt$_3$ or 
Zr$_5$Sb$_3$, the addition of oxygen reduces the $T_c$ value~\cite{Hamamoto2018,Lv2013}. 

In carbon-intercalated Zr$_5$Pt$_3$C$_x$, the $T_c$ value depends 
nonmonotonically on $x$, first increasing up to 7\,K for $x = 0.3$, 
then decreasing to $\sim 4$\,K, as the amount of intercalated C is 
further increased~\cite{Renosto2019,Bhattacharyya2022}.  
The first electronic specific-heat and magnetic penetration-depth studies 
suggested that Zr$_5$Pt$_3$ and Zr$_5$Pt$_3$C$_{0.3}$ are nodal superconductors, implying their unconventional SC 
character~\cite{Renosto2019}. 
However, recent muon-spin rotation and relaxation ($\mu$SR) results are 
consistent with a conventional $s$-wave pairing 
in Zr$_5$Pt$_3$C$_{0.5}$~\cite{Bhattacharyya2022}.
More intriguingly, the theoretical calculations predict Zr$_5$Pt$_3$C$_x$ to be Dirac 
nodal-line semimetals and, as such, good candidates for realizing 
topological SC~\cite{Bhattacharyya2022}. 

Although the superconductivity of Zr$_5$Pt$_3$C$_{x}$ compounds has been investigated
by magnetic- and transport measurements, 
complemented by electronic band-structure calculations, 
the microscopic nature of their SC is still not well established.  
Moreover, the lack of a shared doping makes the conclusions of 
the previous studies regarding the superconducting pairing in 
Zr$_5$Pt$_3$C$_{x}$ inconsistent and hardly comparable~\cite{Renosto2019,Bhattacharyya2022}.
To clarify these issues, we synthesized a series of Zr$_5$Pt$_3$C$_{x}$ 
($x = 0$--0.3) samples, and systematically studied their superconducting properties by means of electri\-cal\ re\-sis\-ti\-vi\-ty,
magnetization, and heat-capacity measurements, complemented by $\mu$SR and nuclear magnetic resonance (NMR) methods. 
We find that Zr$_5$Pt$_3$C$_{x}$ exhibits a fully-gapped superconducting 
state with a preserved time-reversal symmetry. 
Our detailed 
local-probe results suggest a conventional $s$-wave SC
in the Zr$_5$Pt$_3$C$_x$ family, essentially
independent of the C content.  

\section{Experimental details}\enlargethispage{8pt}
\label{sec:Method}
%Polycrystalline Zr$_5$Pt$_3$C$_x$ ($x = 0$--0.3) samples were prepared 
%by the arc-melting method. The resulting
%samples were then annealed under vacuum conditions at 800$^\circ$C over 4 days. 
Polycrystalline Zr$_5$Pt$_3$C$_x$ ($x = 0$--0.3) samples were prepared by arc melting the Zr and C powders and Pt wires with different stoichiometric ratios in a high-purity argon atmosphere. Zr and C powders were firstly mixed and pressed into pellets. The ZrC-pellets and Pt wires were then arc melted. To improve the homogeneity,  the samples were flipped and remelted several times and were then annealed under vacuum conditions at 800$^\circ$C over 4 days.
Room-temperature x-ray powder diffraction
(XRD) measurements were used to check the crystal structure and phase purity of the Zr$_5$Pt$_3$C$_x$ samples, by employing a Shimadzu (XRD-7000) diffractometer. 
The magnetic-susceptibility-, electrical-resistivity-, and heat-capacity measurements were performed using a
Quantum Design MPMS and PPMS 
system. The bulk $\mu$SR measurements were carried out at the multipurpose 
surface-muon spectrometer (Dolly) of the Swiss muon source at the 
Paul Scherrer Institut in Villigen, Switzerland. Both transverse-field (TF) and zero-field (ZF) $\mu$SR measurements were performed. 
As to the former, they allowed us to determine the temperature evolution of the superfluid density. As to the latter, we aimed at searching for a 
possible breaking of time-reversal symmetry in the superconducting state 
of Zr$_5$Pt$_3$C$_x$. 
To exclude the possibility of stray magnetic fields during the ZF-$\mu$SR 
measurements, all the magnets were preliminarily degaussed. 
All the $\mu$SR spectra were collected upon heating and were analyzed 
by means of the \texttt{musrfit} software package~\cite{Suter2012}. 

A series of $^{195}$Pt NMR measurements on powdered 
Zr$_5$Pt$_3$C$_x$ samples in a magnetic field of 4.011\,T provided
the line shapes and the spin-lattice relaxation times. 
We used a continuous-flow CF-1200 cryostat by Oxford Instruments to 
cover the 2 to 300\,K temperature range, with temperatures below 4.2\,K 
being reached
under pumped $^{4}$He conditions. 
The ${}^{195}$Pt reference frequency was determined from the 
${}^{27}$Al resonance, scaled by using the NMR frequency 
tables~\cite{Harris2001}. 
The $^{195}$Pt NMR signal was detected by means of a standard spin-echo 
sequence consisting of $\pi/2$ and $\pi$ pulses of 6 and 12\,$\mu$s, 
with recycling delays ranging from 0.1 to 2\,s in the 2--300\,K
temperature range. 
Spin-lattice relaxation times $T_1$ were measured 
via the inversion-recovery method, using a $\pi$--$\pi/2$--$\pi$ 
pulse sequence with phase cycling for minimizing possible
artifacts. 
The powder samples were used for the x-ray diffraction, magnetization, $\mu$SR, and NMR measurements, 
	while the electrical-resistivity and heat capacity measurements were performed on the cut slabs.

\section{Results and discussion}\enlargethispage{8pt}
\label{sec:results}
\subsection{Crystal structure}\enlargethispage{8pt}
\label{sec:structure}

\begin{figure}[!thp]
	\centering
	\vspace{-6ex}%
	\includegraphics[width=0.47\textwidth,angle=0]{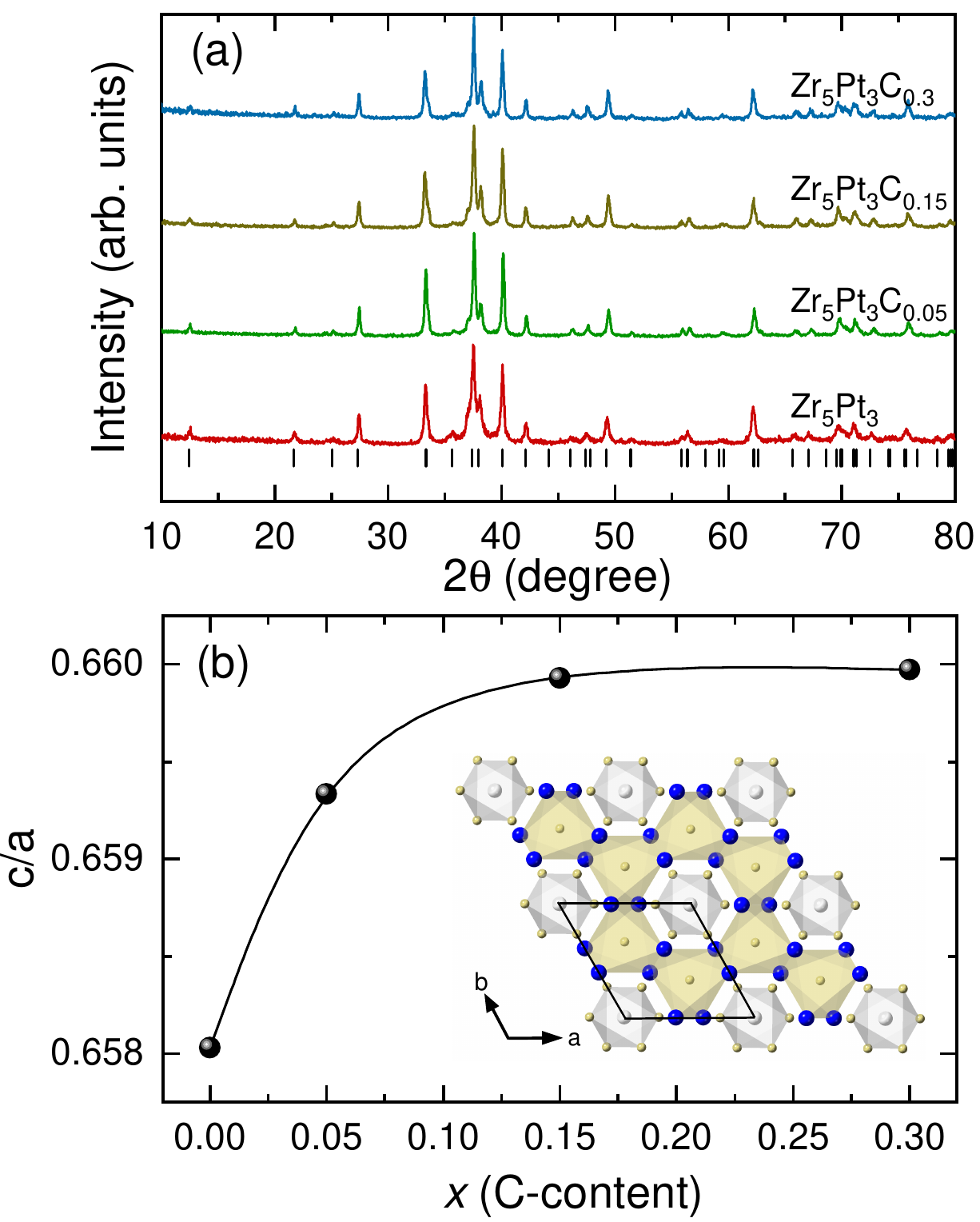}
	\caption{\label{fig:XRD}(a) Room-temperature x-ray powder diffraction 
	patterns for  Zr$_5$Pt$_3$C$_x$ ($x = 0$--0.3). 
		The vertical bars mark the calculated Bragg-peak positions for Zr$_5$Pt$_3$ with a space group $P6_3/mcm$.
		(b) The $c/a$ ratio versus the C-content. The inset shows the crystal structure of Zr$_5$Pt$_3$C viewed along the $c$ axis (solid lines mark
		the unit cell).  Blue, yellow, and gray spheres represent the Pt, Zr, and C atoms, respectively.}
\end{figure}

The crystal structure and the purity of Zr$_5$Pt$_3$C$_x$ ($x=0$--0.3) polycrystalline samples was checked via XRD at room temperature.  
The XRD patterns shown in Fig.~\ref{fig:XRD}(a) 
confirm that all of them share the same 
hexagonal Mn$_5$Si$_3$-type structure, with no discernible traces of foreign phases \tcr{(see details in Fig.~S1 in the
Supplemental Material (SM)~\cite{Supple})}.
As an example, the crystal structure of Zr$_5$Pt$_3$C$_x$ for $x = 1$ 
is shown in the inset of Fig.~\ref{fig:XRD}(b). When $x < 1$, the 
occupation of the 2$b$ sites is less than 1.
We note that, for $x > 0.5$, sizeable amounts of secondary phases 
(mostly ZrPt) appear~\cite{Renosto2019,Bhattacharyya2022}. 
Thus, here we investigate Zr$_5$Pt$_3$C$_x$ samples with $x \le 0.3$.  
The lattice parameters of each sample were obtained by the least-squares method and the results are summarized in Fig~\ref{fig:XRD}(b).
Upon increasing the C-content, $a$ decreases slightly, from 8.182(3)\,\AA\  
(for $x = 0$) to 8.167(4)\,\AA\ (for $x = 0.3$), while $c$ 
increases from 5.384(2)\,\AA\  (for $x = 0$) to 
5.390(3)\,\AA\  (for $x = 0.3$). 
In the Zr$_5$Pt$_3$ parent compound, we obtain lattice parameters 
consistent with previous reports~\cite{Renosto2019,Hamamoto2018}.  
%
%==== figure =============================%
%=== end figure ==========================%
%

However, for the C-intercalated samples, the parameter sets differ.   
In particular, in a previous study, $c$ was found to be almost independent 
of $x$, while $a$ was reported to increase with $x$, resulting in a gradual 
suppression of the $c/a$ ratio~\cite{Renosto2019}.  
In our samples, instead, 
the $c/a$ ratio increases upon increasing 
$x$, to saturate at $x > 0.15$ [see Fig.~\ref{fig:XRD}(b)]. 
Compared to the previous results, our samples show a better crystalline quality, reflected in 
much sharper XRD reflections in Fig.~\ref{fig:XRD}(a).
The reason for such 
discrepant results for
Zr$_5$Pt$_3$C$_x$ is 
not yet clear and requires further investigation.

\subsection{Superconducting temperature 
and lower critical field \texorpdfstring{$H_\mathrm{c1}$}{Hc1}}\enlargethispage{8pt}
\label{sec:Hc1}

%==== figure =============================%
\begin{figure}[!thp]
	\centering
	\vspace{-1ex}%
	\includegraphics[width=0.45\textwidth,angle=0]{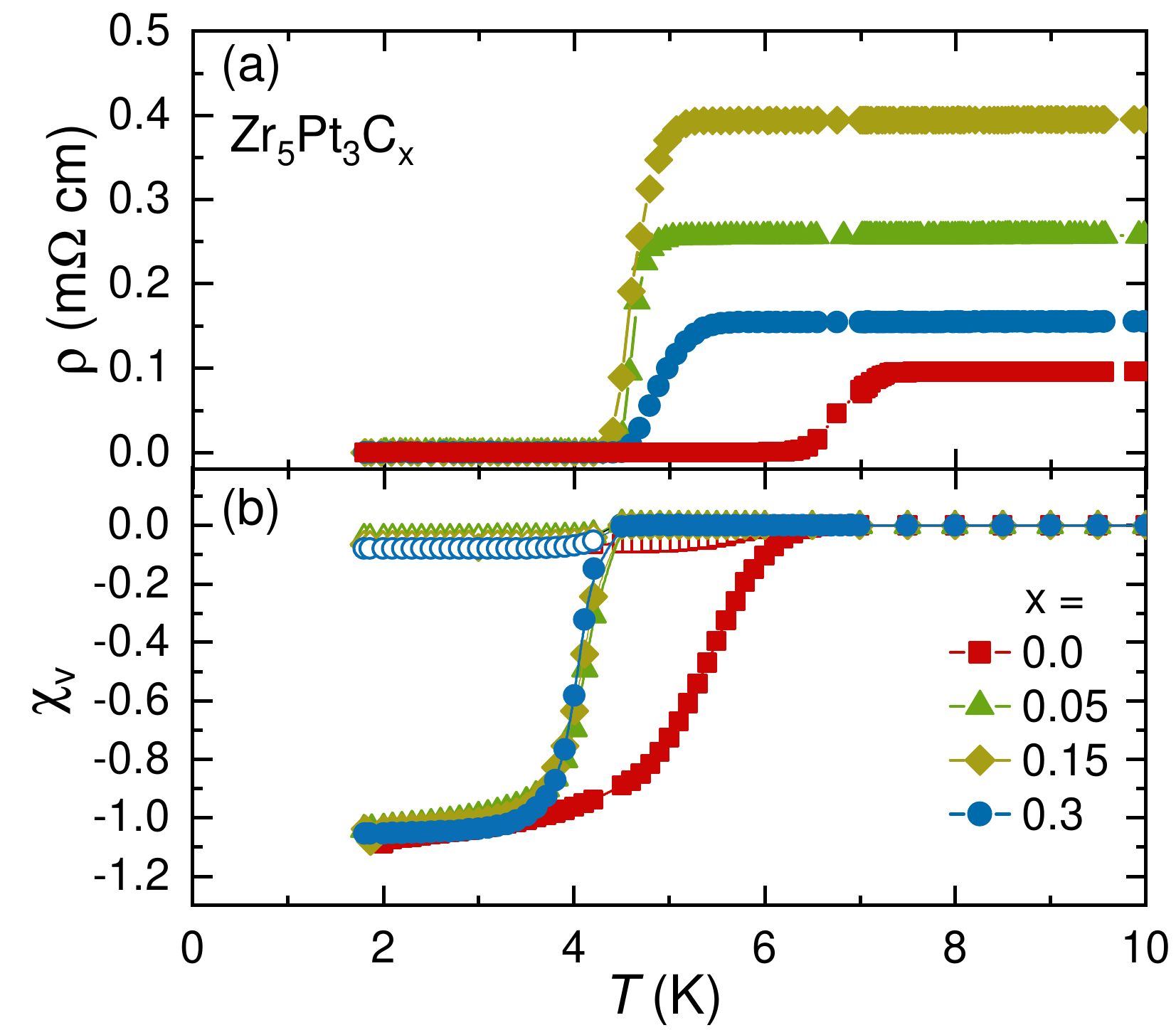}
	\caption{\label{fig:Tc}Temperature-dependent electrical resistivity $\rho(T)$ (a) and magnetic susceptibility $\chi_\mathrm{V}(T)$ (b) for 
	C-intercalated Zr$_5$Pt$_3$.
	While $\rho(T)$ was measured in a zero-field condition, $\chi_\mathrm{V}(T)$ data were collected in a magnetic field of 1\,mT using both the ZFC and FC protocols. 
	The magnetic susceptibilities were corrected by
	considering the demagnetization factor obtained from the field-dependent
	magnetization at 2\,K (base temperature).}
\end{figure}
%=== end figure ==========================%
%

The temperature dependence of the electrical resistivity $\rho(T)$, collected in zero magnetic field, reveals the metallic character of Zr$_5$Pt$_3$C$_x$ ($x$ = 0--0.3). 
The electrical resistivity in the low-$T$ region (below 10\,K) is shown in Fig.~\ref{fig:Tc}(a).  
For Zr$_5$Pt$_3$, the superconducting transition,  with $T_c^\mathrm{onset}$ = 7.2\,K, $T_c^\mathrm{mid}$ = 6.8\,K, and $T_c^\mathrm{zero}$ = 6.1\,K
is clearly visible and the $T_c$s are consistent with previous results~\cite{Renosto2019}. Here $T_c^\mathrm{mid}$ indicates the temperature at the middle of superconducting transition. 
When intercalating carbon into the Zr$_5$Pt$_3$ structure,
$T_c$ is reduced, 
with $T_c^\mathrm{onset}$ = 5.2\,K, $T_c^\mathrm{mid}$ = 4.9\,K, and $T_c^\mathrm{zero}$ = 4.4\,K in 
Zr$_5$Pt$_3$C$_{0.3}$. The slightly different $T_c$ values between the current and the previous work are most likely attributed to a different actual
carbon content~\cite{Renosto2019,Bhattacharyya2022}. 
Since our intercalated samples exhibit almost identical
$T_c$ values with varying C content, our $\mu$SR and NMR measurements 
focused on Zr$_5$Pt$_3$C$_x$ with $x$ = 0 and 0.3.

\begin{figure}[!thp]
	\centering
	\vspace{-1ex}%
	\includegraphics[width=0.47\textwidth,angle=0]{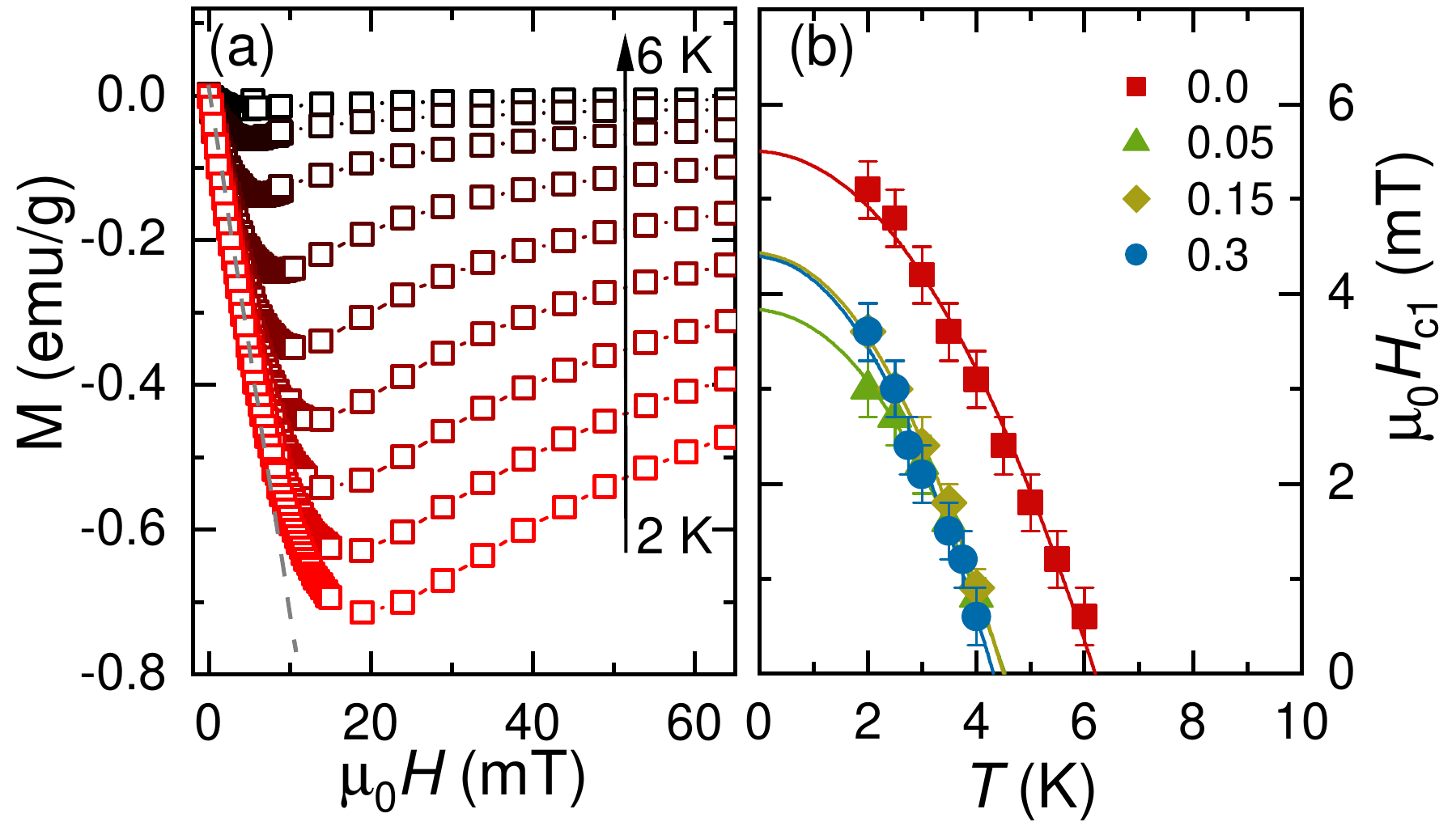}
	\caption{\label{fig:Hc1}(a) Field-dependent magnetization
	$M(H)$ recorded at various temperatures up to $T_c$ for
	Zr$_5$Pt$_3$. 
	(b) Lower critical fields $H_\mathrm{c1}$ vs temperature for Zr$_5$Pt$_3$C$_x$ 
	($x = 0$--0.3). 
	The solid lines represent fits to $\mu_{0}H_{c1}(T) =\mu_{0}H_{c1}(0)[1-(T/T_{c})^2]$.
	For each temperature, the lower critical field $H_\mathrm{c1}$ was determined as the value where $M(H)$ starts deviating from linearity
	(indicated by the dashed line).}
\end{figure}

The superconductivity of Zr$_5$Pt$_3$C$_x$ was further characterized by magnetic susceptibility measurements, using both field-cooled (FC) and zero-field-cooled (ZFC) protocols
in an applied field of 1\,mT. As shown
in Fig.~\ref{fig:Tc}(b), a clear diamagnetic signal appears below the
superconducting transition at $T_c$ = 6.3 and 4.5\,K for Zr$_5$Pt$_3$ and
Zr$_5$Pt$_3$C$_{0.3}$, respectively, in agreement with the values determined from
electrical resistivity in Fig.~\ref{fig:Tc}(a). 
After accounting for the demagnetization
factor, the superconducting shielding fraction of Zr$_5$Pt$_3$C$_x$ samples
is almost 100\%, indicative of bulk SC.
To determine the lower critical field $H_\mathrm{c1}$, essential for performing $\mu$SR measurements
on type-II superconductors, the field-dependent
magnetization $M(H)$ was collected at various temperatures up to $T_c$. As an example, some representative $M(H)$ curves are shown in Fig.~\ref{fig:Hc1}(a) for Zr$_5$Pt$_3$. The C-intercalated samples exhibit 
very similar features.
The resulting $H_\mathrm{c1}$ values as a function of temperature are summarized in Fig.~\ref{fig:Hc1}(b) for Zr$_5$Pt$_3$C$_x$. As shown 
by solid lines, the estimated zero-temperature $H_\mathrm{c1}$ 
values are 
$\mu_0$$H_\mathrm{c1}(0)$ = 5.5(1), 3.8(1), 4.4(1), and 4.4(1)\,mT for $x$ = 0, 0.05, 0.15, and 0.3, respectively. The different $H_\mathrm{c1}$ values of
Zr$_5$Pt$_3$ and Zr$_5$Pt$_3$C$_{0.3}$ are consistent with the magnetic 
penetration depth determined from TF-$\mu$SR measurements (see below). 
%
%==== figure =============================%
%=== end figure ==========================%
%
%
%==== figure =============================%
\begin{figure}[!thp]
	\centering
	%\vspace{-1ex}%
	\includegraphics[width=0.45\textwidth,angle=0]{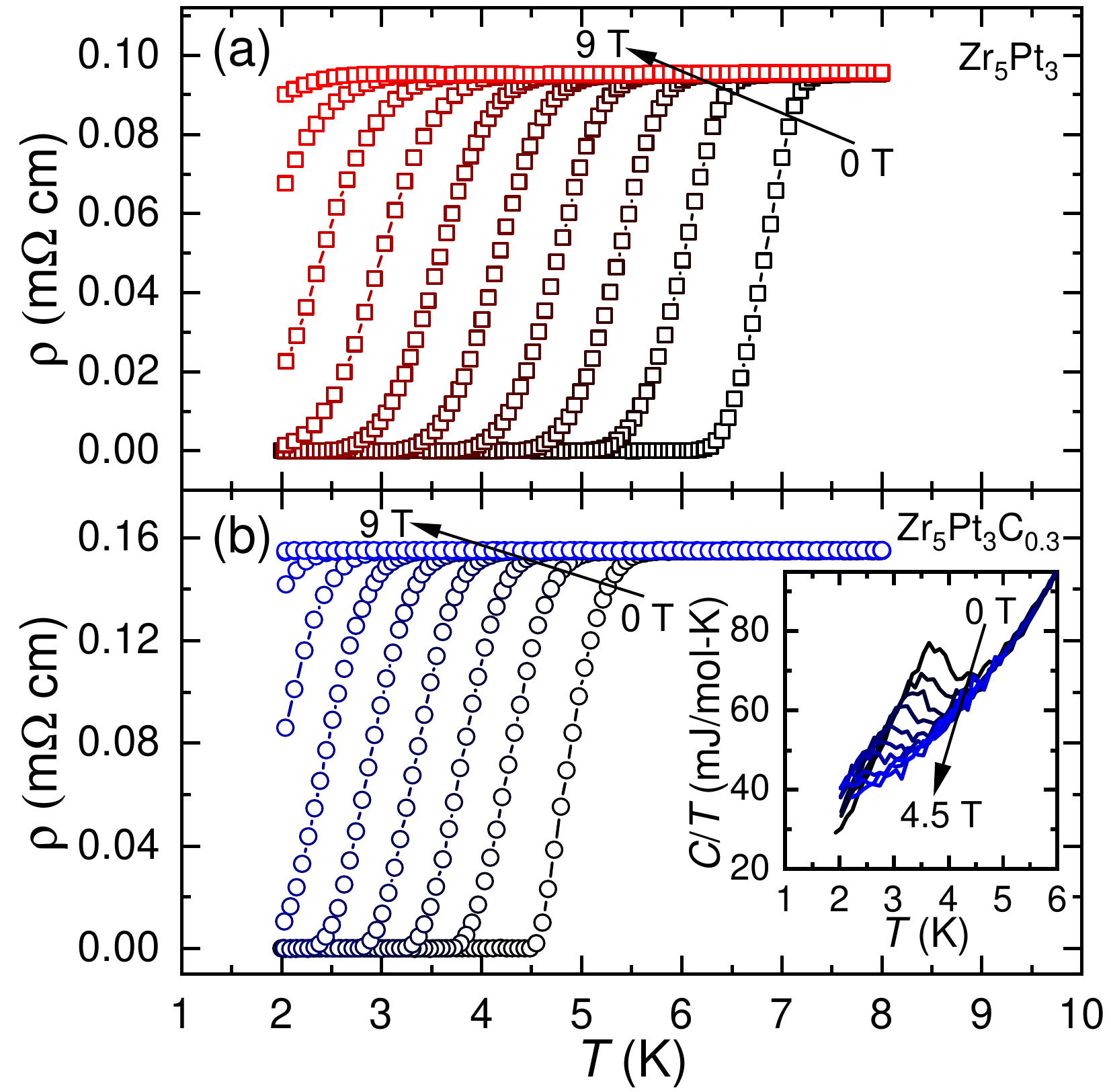}
	\caption{\label{fig:rho}Temperature-dependent electrical resistivity 
	$\rho(T,H)$ collected at various magnetic fields up 9\,T for Zr$_5$Pt$_3$ (a) and Zr$_5$Pt$_3$C$_{0.3}$ (b). 
	Inset of (b): Temperature-dependent specific heat $C(T,H)/T$ collected at various magnetic fields up 4.5\,T. In $\rho(T,H)$ measurements, $T_c$ was determined as the onset of the zero electrical resistivity, i.e., $T_c^\mathrm{zero}$, while in $C(T,H)/T$ measurements, $T_c$ was defined as the midpoint of the superconducting transition.
	The resulting $T_c$ values are summarized in the 
	$H$--$T$ phase diagram shown in Fig~\ref{fig:Hc2}.}
\end{figure}
%=== end figure ==========================%
%

\subsection{Upper critical field \texorpdfstring{$H_\mathrm{c2}$}{Hc2}}\enlargethispage{8pt}
\label{sec:Hc2}

\begin{figure}[!thp]
	\centering
	\vspace{-1ex}%
	\includegraphics[width=0.49\textwidth,angle=0]{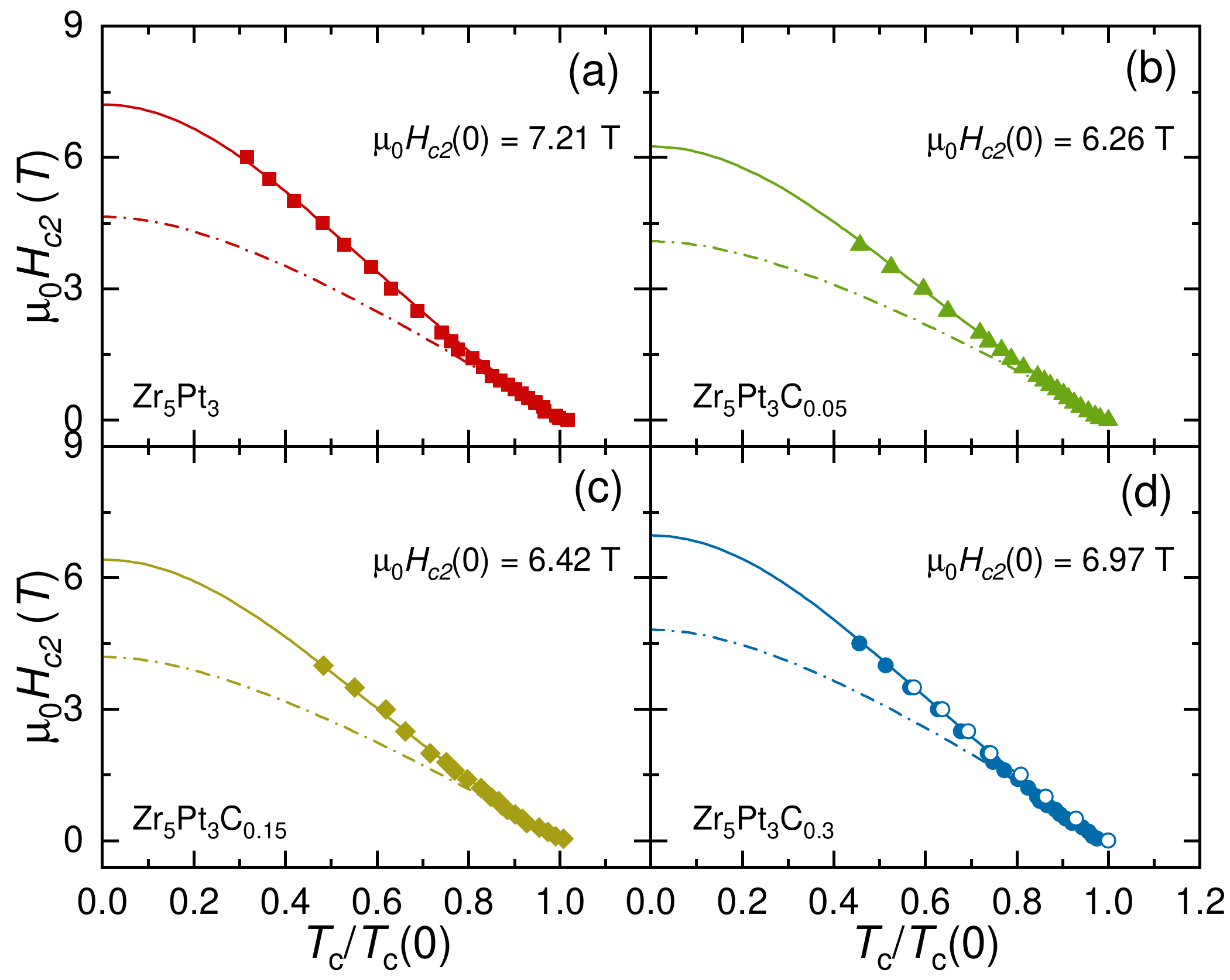}
	\caption{\label{fig:Hc2}Upper critical field $H_\mathrm{c2}$ vs the reduced transition temperature $T_c$/$T_c(0)$ for Zr$_5$Pt$_3$ (a), Zr$_5$Pt$_3$C$_{0.05}$ (b), Zr$_5$Pt$_3$C$_{0.15}$ (c), and Zr$_5$Pt$_3$C$_{0.3}$ (d). The $T_c$ values were determined from measurements shown in Fig.~\ref{fig:rho}. For Zr$_5$Pt$_3$C$_{0.3}$, the $T_c$ values determined from $C(T,H)/T$ (open symbols) are 
	consistent with the values determined from $\rho(T,H)$ (closed symbols). The solid and dash-dotted lines represent fits 
	to the GL- and WHH-models.}
\end{figure}

The upper critical fields $H_\mathrm{c2}$ of Zr$_5$Pt$_3$C$_x$ were 
determined from measurements of the electrical resistivity 
$\rho(T,H)$ in various applied magnetic fields.
As an example, the $\rho(T,H)$ curves of Zr$_5$Pt$_3$ and Zr$_5$Pt$_3$C$_{0.3}$ are shown in Figs.~\ref{fig:rho}(a) and \ref{fig:rho}(b), respectively. 
In an
applied field, the superconducting transition shifts toward lower 
temperatures and broadens. For Zr$_5$Pt$_3$C$_{0.3}$, as shown in the inset of Fig.~\ref{fig:rho}(b), also specific-heat 
measurements in
various magnetic fields 
were performed. 
Since, for $x = 0.3$, the $T_c$ values determined from the specific 
heat $C(T,H)/T$ coincide
with $T_c^\mathrm{zero}$ determined from the electrical-resistivity measurements [see Fig.~\ref{fig:Hc2}(d)],
we used the $T_c^\mathrm{zero}$ values as a criterion to determine 
$H_\mathrm{c2}(0)$ for all the Zr$_5$Pt$_3$C$_x$ samples. 
The $H_\mathrm{c2}(T)$ vs $T_c/T_c(0)$ data [here, $T_c(0)$ is the 
transition temperature in zero field] 
are summarized in Fig.~\ref{fig:Hc2}.
Each $H_\mathrm{c2}(T)$ curve was analyzed by means of 
Ginzburg–Landau (GL), $H_{c2} = H_{c2}(0)(1-t^2)/(1+t^2)$,~\cite{Zhu2008} 
and Werthamer–Helfand–Hohenberg 
(WHH) models~\cite{Werthamer1966}.
As shown by the dash-dotted lines,
the WHH model can describe the $H_\mathrm{c2}(T)$ data 
reasonably well up to 2\,T. 
However, at higher magnetic fields, this model deviates significantly from the experimental data and provides underestimated $H_\mathrm{c2}$ values.
By contrast, as shown by the solid lines in Fig.~\ref{fig:Hc2}, the GL model 
agrees remarkably well with the experimental
data and provides $\mu_0H_\mathrm{c2}(0)$ = 7.21(5), 6.26(4), 6.42(9), and 6.97(6)\,T for $x$ = 0, 0.05, 0.15, and 0.3, respectively.

%==== figure =============================%
%=== end figure ==========================%
%

In the GL theory of superconductivity, the magnetic penetration 
depth $\lambda$ is related to the coherence length $\xi$, and the lower 
critical field via $\mu_{0}H_\mathrm{c1} = (\Phi_0 /4 \pi \lambda^2)[$ln$(\kappa)+\alpha(\kappa)]$, 
where $\Phi_0 = 2.07 \times 10^{-3}$\,T~$\mu$m$^{2}$ is the quantum of 
magnetic flux, $\kappa$ = $\lambda$/$\xi$ is the GL parameter, 
and $\alpha(\kappa)$ is a parameter which converges to 0.497 for $\kappa$ $\gg$ 1~\cite{Brandt2003}. 
By using $\mu_{0}H_\mathrm{c1}$ and $\xi$ values [calculated from $\mu_{0}H_\mathrm{c2}(0)$ = $\Phi_0$/2$\pi\xi(0)^2$], the resulting $\lambda_\mathrm{GL}$ =  366(4) and 415(6) for Zr$_5$Pt$_3$ and  Zr$_5$Pt$_3$C$_{0.3}$ are
compatible with the experimental value determined from $\mu$SR 
data. All the superconducting parameters are summarized in Table~\ref{tab:parameter}.
A GL parameter $\kappa \gg 1$ confirms again that Zr$_5$Pt$_3$C$_x$  are type-II superconductors.

\subsection{\texorpdfstring{$\mu$SR}{muSR} study}\enlargethispage{8pt}
\label{sec:musr}
\subsubsection{Transverse-field \texorpdfstring{$\mu$SR}{muSR}} %\enlargethispage{8pt}
\begin{figure}[!thp]
	\centering
	\vspace{-1ex}%
	\includegraphics[width=0.49\textwidth,angle=0]{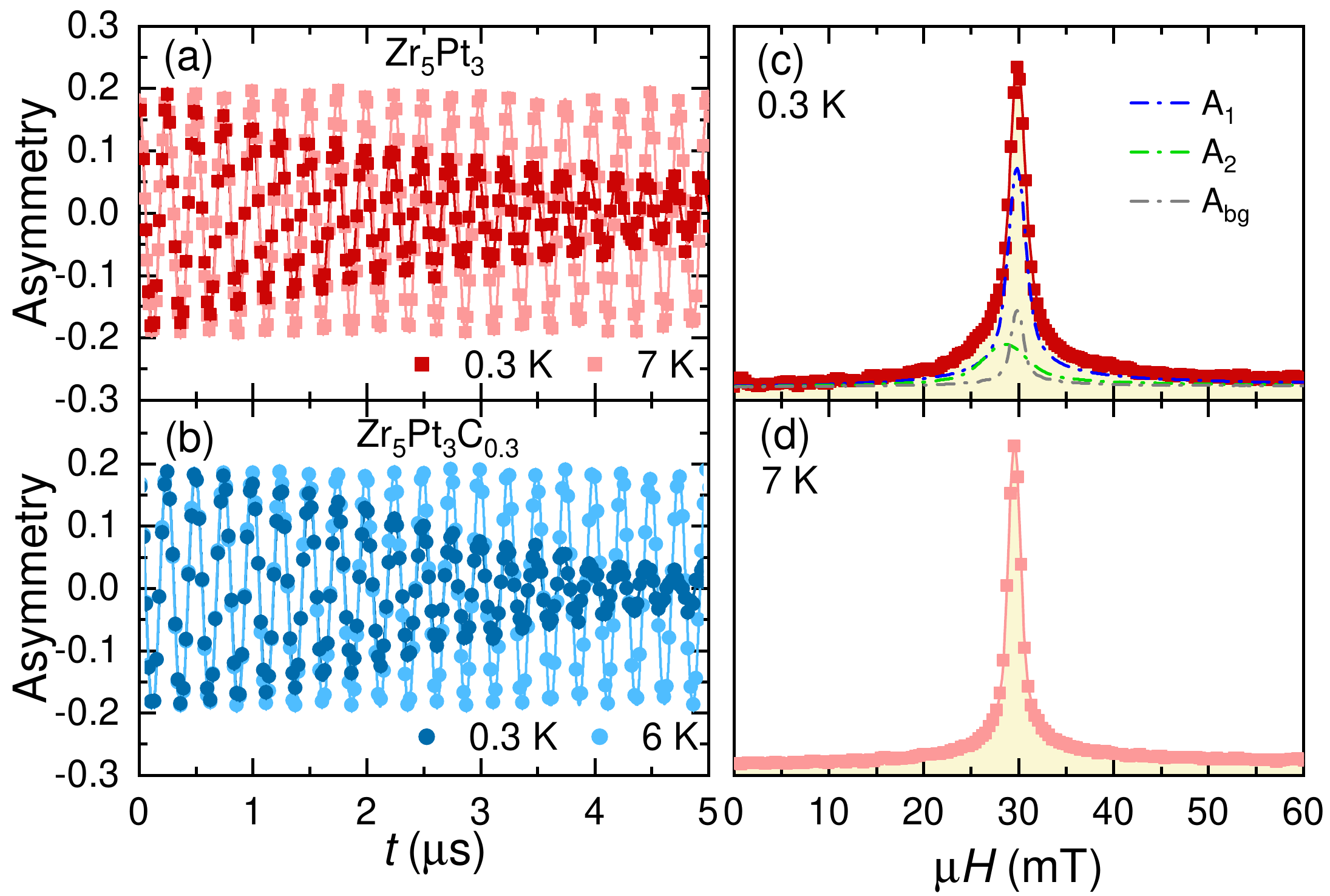}
	\caption{\label{fig:TF_muSR}%
	TF-$\mu$SR spectra collected in the normal and superconducting 
	states in an applied magnetic field of 30\,mT for Zr$_5$Pt$_3$ (a) 
	and Zr$_5$Pt$_3$C$_{0.3}$ (b). Fast Fourier transforms of the TF-$\mu$SR data shown in (a) 
	at 0.3\,K (c) and at 7\,K (d).  
	For Zr$_5$Pt$_3$, solid lines are fits to Eq.~\eqref{eq:TF_muSR} using two oscillations, which are also shown separately as
	dash-dotted lines in (c), together with a background contribution. For Zr$_5$Pt$_3$C$_{0.3}$, 
	solid lines are fits to Eq.~\eqref{eq:TF_muSR} with a single oscillation. Note the clear field-distribution broadening due to FLL below $T_c$. 	
	}
\end{figure}

\label{ssec:TF}
To investigate 
Zr$_5$Pt$_3$ and Zr$_5$Pt$_3$C$_{0.3}$
at a microscopic level, we carried out systematic TF-$\mu$SR
measurements in an applied field of 30\,mT, i.e., more than twice their $H_\mathrm{c1}$ values [see Fig.~\ref{fig:Hc1}(b)].
Representative TF-$\mu$SR spectra collected in the superconducting and normal
states of Zr$_5$Pt$_3$ and Zr$_5$Pt$_3$C$_{0.3}$ are shown in Figs.~\ref{fig:TF_muSR}(a) and \ref{fig:TF_muSR}(b), respectively.
For both compounds, the normal-state spectra show
essentially no damping, thus reflecting the uniform 
field distribution, as well as the lack of magnetic impurities.  
In the superconducting state (below $T_c$), instead, the significantly enhanced damping reflects the inhomogeneous field distribution due to 
the development of a flux-line lattice (FLL)~\cite{Yaouanc2011,Amato1997,Blundell1999,Maisuradze2009}. The broadening of the field distribution is clearly visible in 
%
%==== figure =============================%
%=== end figure ==========================%
%
%
Figs.~\ref{fig:TF_muSR}(c) and \ref{fig:TF_muSR}(d), where the fast-Fourier transform (FFT) spectra of the corresponding TF-$\mu$SR data of Zr$_5$Pt$_3$ are presented. 
To describe the asymmetric field distribution [e.g., see FFT at 0.3\,K in Fig.~\ref{fig:TF_muSR}(c)], the TF-$\mu$SR spectra were modeled using
\begin{equation}
	\label{eq:TF_muSR}
	A_\mathrm{TF}(t) = \sum\limits_{i=1}^n A_i \cos(\gamma_{\mu} B_i t + \phi) e^{- \sigma_i^2 t^2/2} +
	A_\mathrm{bg} \cos(\gamma_{\mu} B_\mathrm{bg} t + \phi).
\end{equation}

\begin{figure}[!thp]
	\centering
	\vspace{-1ex}%
	\includegraphics[width=0.45\textwidth,angle=0]{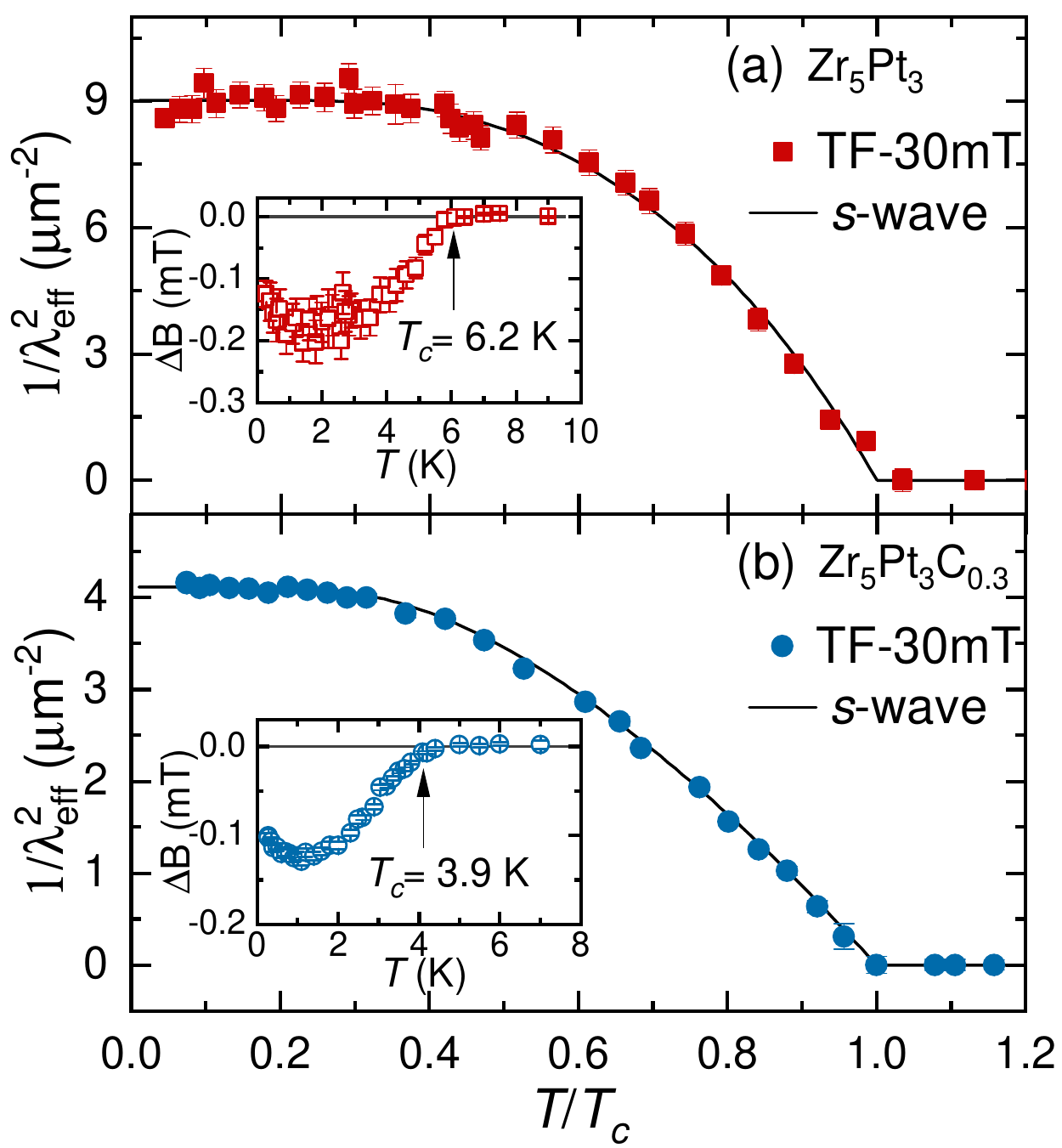}
	\caption{\label{fig:superfluid}%
     Superfluid density vs reduced temperature $T/T_c$ for Zr$_5$Pt$_3$ 
	(a) and Zr$_5$Pt$_3$C$_{0.3}$ (b). Insets show the diamagnetic shift
	$\Delta$$B(T)$. The solid lines in the main panels
	represent fits to a fully gapped $s$-wave model with a single energy gap.}
\end{figure}

\noindent %
Here $A_{i}$, $A_\mathrm{bg}$ and $B_{i}$, $B_\mathrm{bg}$ 
are the initial asymmetries and local fields sensed by implanted muons in the 
sample and sample holder (copper, which normally shows zero muon-spin depolarization), $\gamma_{\mu}$/2$\pi$ = 135.53\,MHz/T 
is the muon gyromagnetic ratio, $\phi$ is a shared initial phase, and $\sigma_{i}$ 
is the Gaussian relaxation rate of the $i$th component. 
Generally, the field distribution in the superconducting state is material dependent: the more asymmetric it is, the more components are required to describe it. 
Here, we find that, while two oscillations (i.e., $n = 2$) are required to properly describe the TF-$\mu$SR spectra of Zr$_5$Pt$_3$, 
a single oscillation is sufficient for
Zr$_5$Pt$_3$C$_{0.3}$.
For Zr$_5$Pt$_3$, the dash-dotted lines in Fig.~\ref{fig:TF_muSR}(c) 
represent the two
components at 0.3\,K ($A_1$ and $A_2$) and the background signal ($A_\mathrm{bg}$).
Above $T_c$, the muon-spin relaxation rate is small and temperature-independent, but below $T_c$ it starts to increase due to the onset of FLL and the increased superfluid density.
At the same time, a diamagnetic field shift, $\Delta B (T) = \langle B \rangle - B_\mathrm{appl}$, appears below $T_c$,  
with $\langle B \rangle = (A_1\,B_1 + A_2\,B_2)/A_\mathrm{tot}$, 
$A_\mathrm{tot} = A_1 + A_2$, and $B_\mathrm{appl} = 30$\,mT (see insets 
in Fig.~\ref{fig:superfluid}). The effective Gaussian relaxation rate can be calculated
from $\sigma_\mathrm{eff}^2/\gamma_\mu^2 = \sum_{i=1}^2 A_i [\sigma_i^2/\gamma_{\mu}^2 - \left(B_i - \langle B \rangle\right)^2]/A_\mathrm{tot}$~\cite{Maisuradze2009}.
Then, the superconducting Gaussian relaxation rate $\sigma_\mathrm{sc}$, can be extracted by subtracting 
the nuclear contribution according to $\sigma_\mathrm{sc} = \sqrt{\sigma_\mathrm{eff}^{2} - \sigma^{2}_\mathrm{n}}$. 
Here, $\sigma_\mathrm{n}$ is
the nuclear relaxation rate,
almost constant in the covered 
temperature range and extremely small for Zr$_5$Pt$_3$C$_x$, 
as confirmed also by ZF-$\mu$SR data (see Fig.~\ref{fig:ZF_muSR}). 
At low magnetic fields ($H_\mathrm{appl} / H_\mathrm{c2} \sim$ 0.004 $\ll$\,1), the effective magnetic penetration depth $\lambda_\mathrm{eff}$ and, thus, the superfluid density $\rho_\mathrm{sc}$ ($\propto  \lambda_\mathrm{eff}^{-2}$), can be calculated using
$\sigma_\mathrm{sc}^2(T)/\gamma^2_{\mu} = 0.00371\Phi_0^2/\lambda_\mathrm{eff}^4(T)$~\cite{Barford1988,Brandt2003}.

The superfluid density $\rho_\mathrm{sc}$ of Zr$_5$Pt$_3$ and 
Zr$_5$Pt$_3$C$_{0.3}$ vs the reduced temperature 
$T/T_c$ is shown in Figs.~\ref{fig:superfluid}(a) and \ref{fig:superfluid}(b), respectively.
%
%==== figure =============================%
%=== end figure ==========================%
%
In both cases, the temperature-invariant superfluid density below $T_c$/3
suggests the absence of low-energy excitations and, hence, a fully-gapped
superconducting state. Consequently, the $\rho_\mathrm{sc}(T)$ was analyzed
by means of a fully-gapped $s$-wave model:
\begin{equation}
	\label{eq:rhos}
	\rho_\mathrm{sc}(T) = \frac{\lambda_\mathrm{eff}^{-2}(T)}{\lambda_0^{-2}} = 1 + 2\int^{\infty}_{\Delta(T)} \frac{\partial f}{\partial E} \frac{EdE}{\sqrt{E^2-\Delta^2(T)}}.
\end{equation}
Here, $f = (1+e^{E/k_\mathrm{B}T})^{-1}$ is the Fermi function; $\Delta(T)$ is the  su\-per\-con\-duc\-ting\--gap function, assumed to follow $\Delta(T) = \Delta_0 \mathrm{tanh} \{1.82[1.018(T_\mathrm{c}/T-1)]^{0.51} \}$~\cite{Tinkham1996,Carrington2003}; $\lambda_0$ and $\Delta_0$ are the magnetic penetration depth and the superconducting gap at 0\,K, respectively.  
As shown by the solid lines in Fig.~\ref{fig:superfluid}, the $s$-wave 
model describes $\rho_\mathrm{sc}(T)$ very well across the entire
temperature range with the fit parameters: $\Delta_0$ = 1.20(2) and 0.60(2)\,meV, and $\lambda_0$ = 333(3) and 493(2)\,nm
for Zr$_5$Pt$_3$ and Zr$_5$Pt$_3$C$_{0.3}$, respectively. 
The gap value and the magnetic penetration depth of 
Zr$_5$Pt$_3$C$_{0.3}$ are close to those of
Zr$_5$Pt$_3$C$_{0.5}$~\cite{Bhattacharyya2022}.
We considered also the dirty-limit model~\cite{Tinkham1996}, 
which turned out to describe fairly well the $\rho_\mathrm{sc}(T)$ 
of Zr$_5$Pt$_3$ and Zr$_5$Pt$_3$C$_{0.3}$, yielding slighly lower 
superconducting gap values (see details in Table~\ref{tab:parameter}).

\subsubsection{Zero-field \texorpdfstring{$\mu$SR}{muSR}} %\enlargethispage{8pt}
\label{ssec:ZF}

\begin{figure}[!thp]
	\centering
	\vspace{-3ex}%
	\includegraphics[width=0.47\textwidth,angle=0]{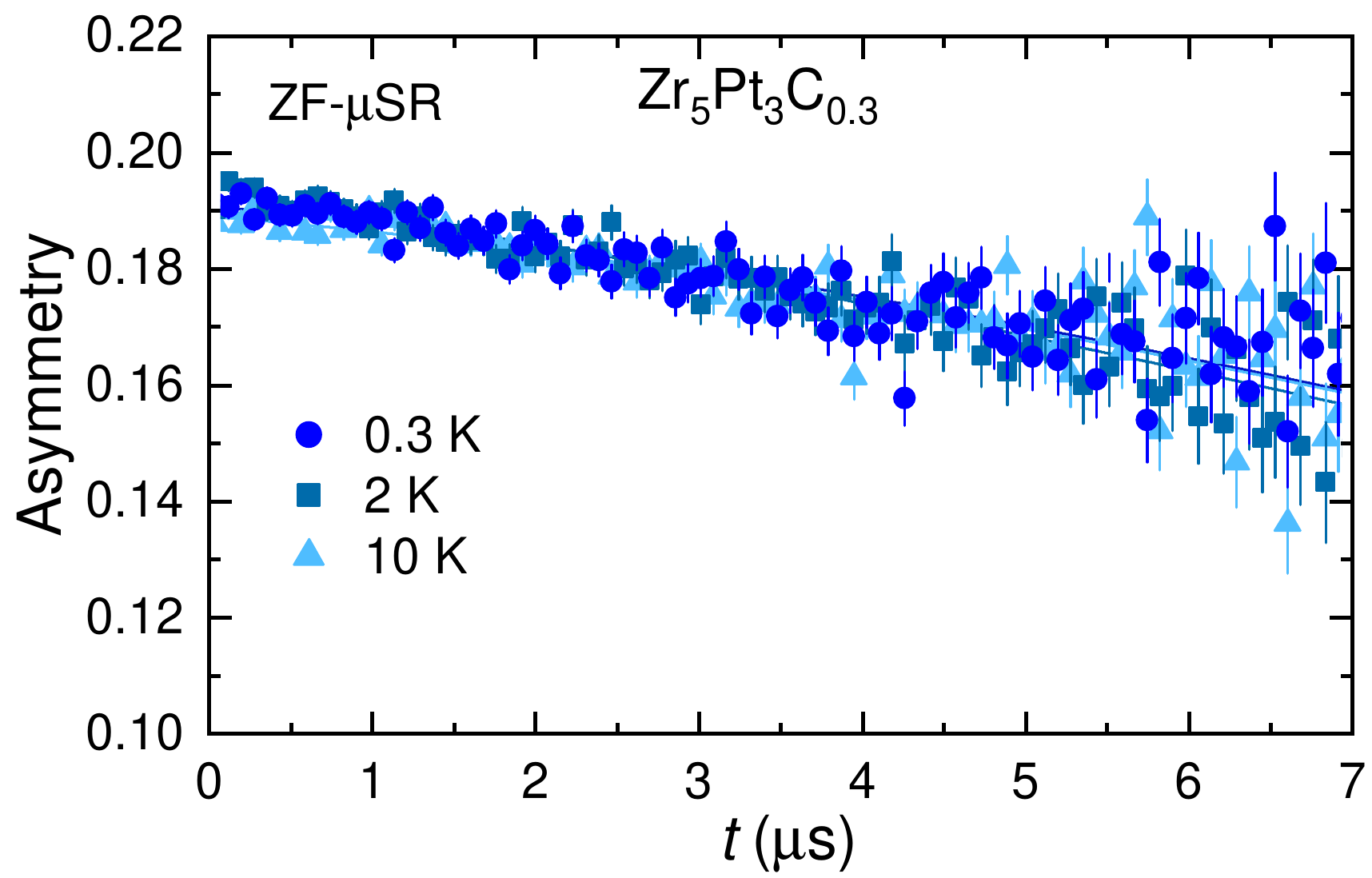}
	\caption{\label{fig:ZF_muSR}%
	Representative ZF-$\mu$SR spectra in the normal (10\,K) and the superconducting state (0.3 and 2\,K) of Zr$_5$Pt$_3$C$_{0.3}$. 
	Solid lines are fits to the equation described in the text. None of the datasets shows noticeable changes with temperature, suggesting the absence of spontaneous magnetic fields in the superconducting state.}
\end{figure}

To verify the possible breaking of time-reversal symmetry in 
Zr$_5$Pt$_3$C$_{0.3}$, we also performed ZF-$\mu$SR measurements 
in both its normal- and superconducting states. As shown in Fig.~\ref{fig:ZF_muSR}, 
neither coherent oscillations nor fast decays could be identified in 
the spectra collected above (10\,K) and below $T_c$ (2\,K and 0.3\,K), hence 
implying the lack of any magnetic order or fluctuations. 
The weak muon-spin relaxation in absence of an external magnetic 
field is mainly due to the randomly oriented nuclear moments, 
which can be modeled by a Gaussian Kubo-Toyabe relaxation function, 
$G_\mathrm{KT} = [\frac{1}{3} + \frac{2}{3}(1 -\sigma_\mathrm{ZF}^{2}t^{2})\,\mathrm{e}^{-\sigma_\mathrm{ZF}^{2}t^{2}/2}]$
~\cite{Kubo1967,Yaouanc2011}.
%
%==== figure =============================%
%=== end figure ==========================%
%
Here, $\sigma_\mathrm{ZF}$ is the zero-field Gaussian relaxation rate. 
The solid lines in Fig.~\ref{fig:ZF_muSR} represent fits to 
the data by considering also an additional zero-field Lorentzian 
relaxation $\Lambda_\mathrm{ZF}$, i.e., $A_\mathrm{ZF}(t) = A_\mathrm{s} G_\mathrm{KT} \mathrm{e}^{-\Lambda_\mathrm{ZF} t} + A_\mathrm{bg}$. 
The relaxation rates in the normal- and the superconducting states are almost 
identical, as confirmed by the practically overlapping ZF-$\mu$SR 
spectra above and below $T_c$. 
The resulting relaxations at 0.3\,K are listed in Table~\ref{tab:parameter}. 
This lack of evidence for an additional 
$\mu$SR relaxation below $T_c$ excludes a possible time-reversal 
symmetry breaking in the superconducting state of Zr$_5$Pt$_3$C$_{0.3}$.

\subsection{\texorpdfstring{${}^{195}$P\MakeLowercase{t}}{195Pt}-NMR study}\enlargethispage{8pt}
\label{sec:nmr}

Zr$_5$Pt$_3$C$_x$ samples contain three NMR-active nuclei, 
i.e., ${}^{13}$C, ${}^{91}$Zr, and ${}^{195}$Pt.
However, the 1.1\% isotopic abundance of ${}^{13}$C, together with  
the low carbon content of the samples and its omnipresence in the 
probehead, made it difficult to use the ${}^{13}$C NMR signal of 
Zr$_5$Pt$_3$C$_x$.  
The low-frequency, low-abundance ${}^{91}$Zr, associated with 
its quadrupole effects ($I = \nicefrac{5}{2}$) made it unsuitable, too. 
Consequently, in the current work, we focus %only 
on the ${}^{195}$Pt NMR results. \tcr{A representative full-frequency
scan is shown in Fig.~S2, while the frequency ranges covered at low-$T$
are listed in Table~S1 of SM~\cite{Supple}.} %in the Supplemental Material~\cite{Supple}.}

\subsubsection{Static electronic properties: Knight shift}%\enlargethispage{8pt}
\label{sec:nmr_shifts}

\begin{figure}[ht]
	\centering
	%\vspace{-1ex}
	\includegraphics[width=0.48\textwidth,angle=0]{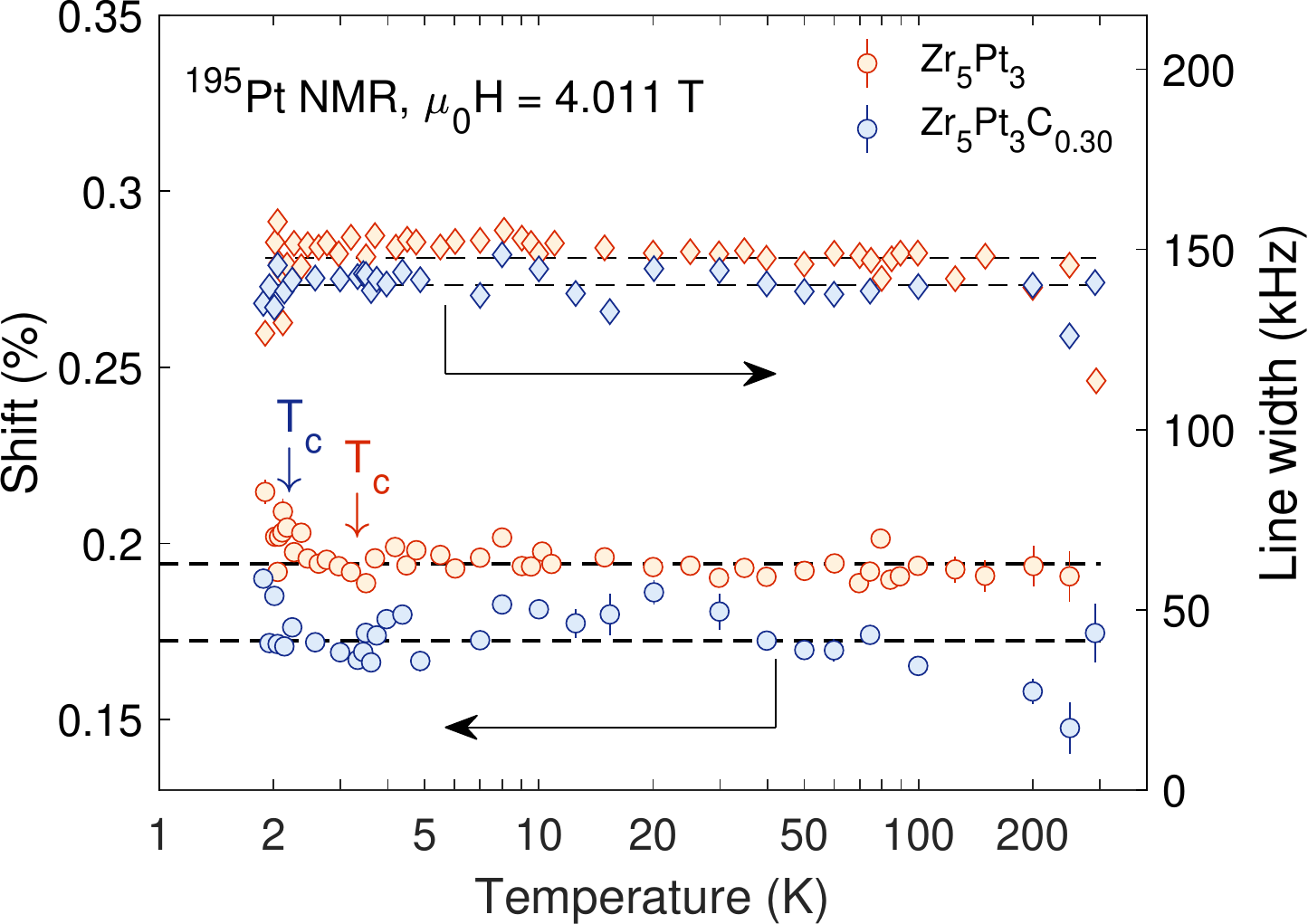}
	\caption{\label{fig:zr5pt3_shift}${}^{195}$Pt NMR frequency shift
		(left scale) and linewidth (right scale) as function
		of temperature. In the normal state, both parameters are constant
		with temperature. Below $T_c$, we observe an increase in shift 
		(see text for details). The arrows mark the $T_c$ values.} 
		%of Zr$_5$Pt$_3$.
\end{figure}

The ${}^{195}$Pt NMR lineshapes (\tcr{see Fig.~S3 in SM~\cite{Supple}}) %Supplemental Material~\cite{Supple}})
were recorded by a standard spin-echo 
sequence and successively fitted with the \texttt{Dmfit} 
software~\cite{Massiot2002}. 
By assuming a purely Gaussian profile for the 
spin-\nicefrac{1}{2} nuclei, we 
determined the precise peak positions and line widths. 
The Pt atoms reside in $6g$ sites~\cite{Renosto2019} and are bonded 
in a 9-coordinate geometry to Zr atoms. Since the latter occupy two 
inequivalent sites ($6g$ and $4d$), of which the first with a 9\% 
spread in Pt-Zr distances, relatively wide $^{195}$Pt NMR lines 
are expected. Due to the large width of the lines, the shifts appear as 
small (\tcr{see Fig.~S3~\cite{Supple}}). Yet, numerical fits provide shift values 
typical of metallic compounds (i.e., a fraction of percent). 
As shown in Fig.~\ref{fig:zr5pt3_shift}, in the normal state, both 
Zr$_5$Pt$_3$ and Zr$_5$Pt$_3$C$_{0.3}$ exhibit a 
temperature-independent frequency shift (known as Knight shift, $K$) 
and li\-ne\-width (full width at half maximum, $\varGamma_{\mathrm{FWHM}}$).  
In Zr$_5$Pt$_3$ (red symbols), 
%
%==== figure =============================%
%=== end figure ==========================%
%
we find $K = 0.195$\% and $\varGamma_{\mathrm{FWHM}} = 150$\,kHz, while 
in Zr$_5$Pt$_3$C$_{0.3}$ (blue symbols) these values are 10\% and 5\% lower, respectively.

Similar features (i.e., wide lines, small and/or positive shifts, and no 
temperature dependence) have been reported also in NiPtP metallic 
glasses~\cite{Hines1978} or in Pt-Mo alloys~\cite{Weisman1967}.
These are to be compared with the $-3.5$\% $^{195}$Pt shift in 
elemental metallic Pt, 
whose large negative value is known to arise from a dominant 
core-polarization contribution~\cite{Yafet1964}. 
In Mott's two-band ($s + d$) model, the density of 
states of the narrow $d$-band is much greater than that of the much 
wider $s$-band. Hence, any property which depends on the 
electronic density of 
states will, in general, be dominated by the unfilled $d$-band.
In our case, the resulting positive increase in $^{195}$Pt Knight shift  
provides evidence that the Pt $d$-states are filled, 
consistent with a charge transfer from Zr to the transition-metal 
atoms (we recall that Pt is twice as electronegative as Zr on the Pauling scale). 
In addition, the wide $^{195}$Pt NMR linewidths, besides 
a distribution of atomic environments, suggest also a 
distribution in the degree of the $d$-state filling.
In our case, it is precisely this type of filling which lowers 
the $d$ character (and reduces the orbital effects), hence accounting 
for the positive NMR shift we observe. Of course, for quantitative 
results one has to consider also the spin-orbit, exchange correlation, 
and $s$-$d$ hybridization effects~\cite{Carter1970,Carter1977}.

%
%==== figure =============================%
\begin{figure}[th!]
	\centering
	%\vspace{-1ex}%
	\includegraphics[width=0.43\textwidth,angle=0]{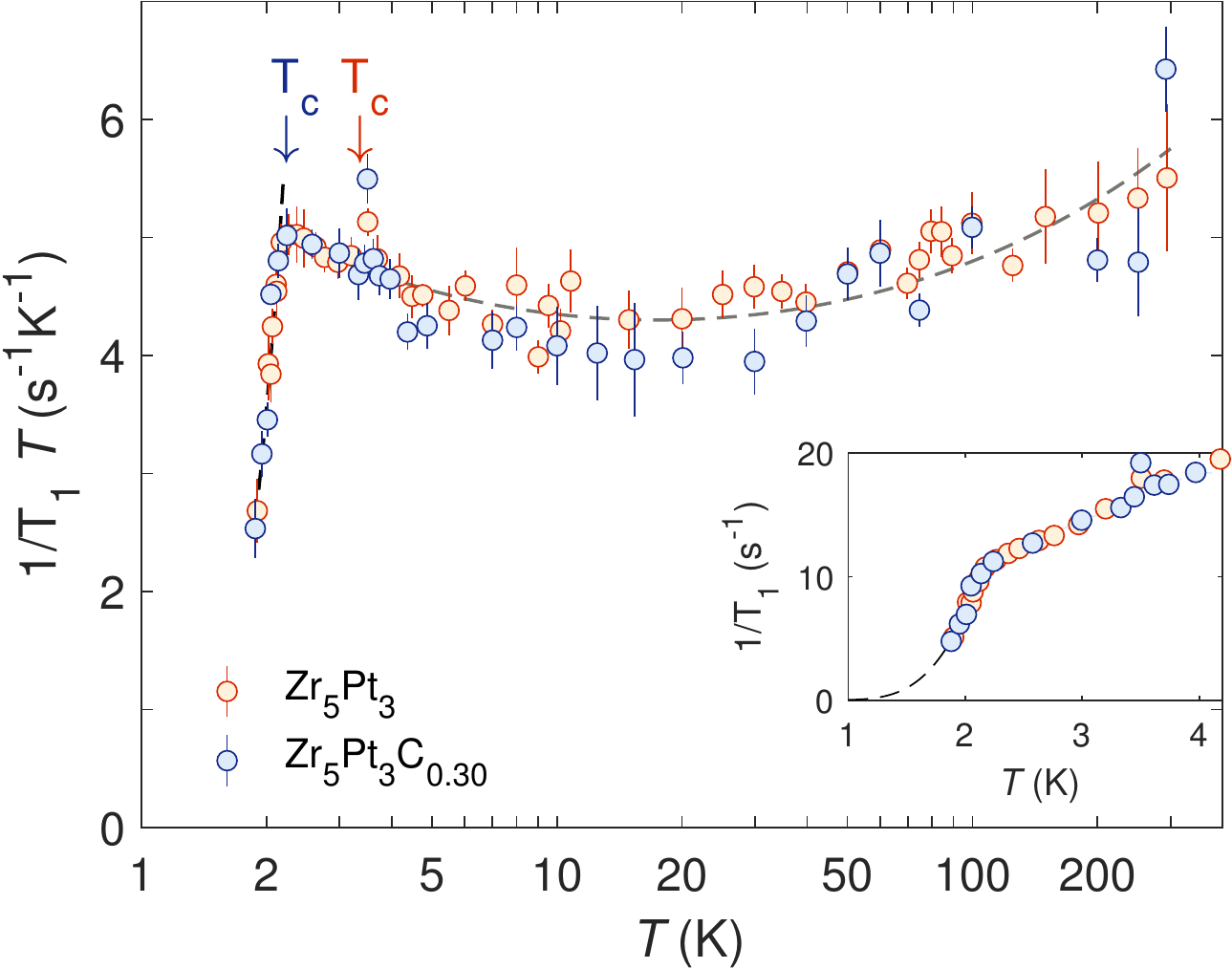}
	\caption{\label{fig:1_t1t}$(T_1T)^{-1}$ vs.\ $T$ for the pure and 
	C-doped	\ZR\ samples. Two fits are proposed: normal state 
	(gray dashed line) and superconducting state (black dashed line). 
	Inset: 
	%$1/(T_1T)$ for $T \leq 8$\,K; 
	${T_1}^{-1}$ for $T \leq 4.2$\,K;
	Fit of the thermally-activated 
	relaxation rate (black dashed line).
	The initial unchanged trend below $T_{c}$ might be due to the 
	relaxation of nuclei inside the vortex cores.
	The arrows mark the $T_c$ values determined from NMR shift in Fig.~\ref{fig:zr5pt3_shift}. 
	The details 
	of $T_1$ measurements \tcr{and their analysis can be found in Figs.~S4-S5
	of SM~\cite{Supple}}.} %in the Supplementary Materials~\cite{Supple}}.}
\end{figure}
%=== end figure ==========================%
%

The orbital Knight shift, here preponderant and temperature independent, 
not only justifies our normal-state results, but also explains why 
the change in shift in the superconducting phase is so small. 
The increase in shift with decreasing temperature, we observe below $T_c$,
is surprising. A possible reason for this might be the
locally higher magnetic field at the SC vortex cores (see further).
A similar behavior has been reported, e.g., in V$_3$Ga~\cite{Clogston1964},
where it was attributed to a significant reduction of the spin susceptibility.
Independently of sign, a change in $K$ across 
$T_c$ represents a clear fingerprint of a fully-gapped conventional 
SC in Zr$_5$Pt$_3$C$_x$.  

\subsubsection{Dynamic electronic properties: Relaxation rates}
\label{sec:nmr_relax}

\begin{figure}[h]
	\centering
	\includegraphics[width=0.45\textwidth,angle=0]{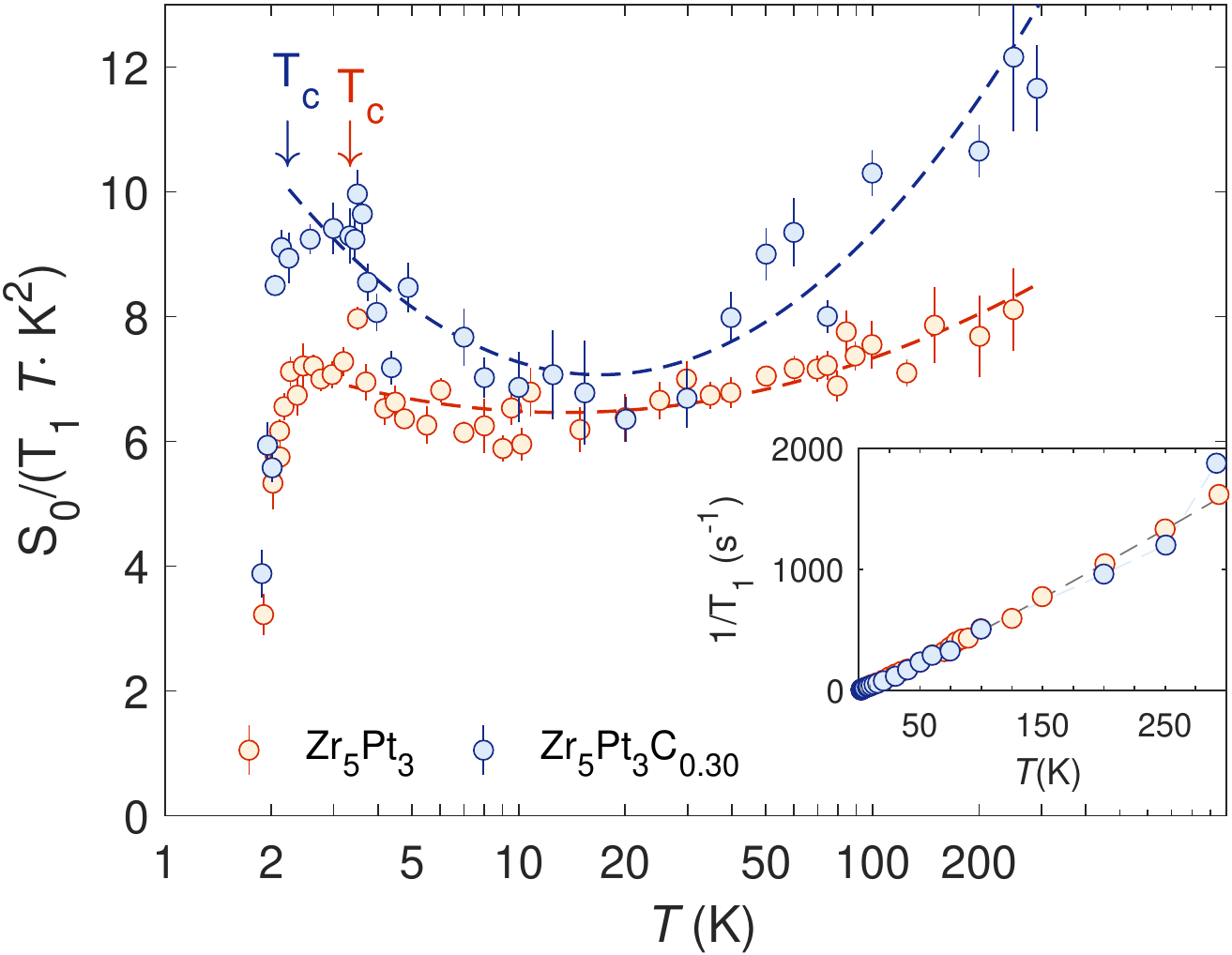}
	\vspace{-2ex}%
	\caption{\label{fig:korringa}The Korringa ratio $S_0/S$ %coefficient 
	(expressed as $1/\alpha$) is always above unity. Here, %$\alpha < 1$  
	$S_0/S > 1$ indicates the presence %of antiferromagnetic 
	non-$s$-type hyperfine interactions. %yet low} electronic correlations.
	Inset: the spin-lattice relaxation rate. 
	The arrows mark the $T_c$ values determined from NMR shift in Fig.~\ref{fig:zr5pt3_shift}.}
	%The arrow marks the 
	%$T_c$ values of \tcr{pure and C-doped} Zr$_5$Pt$_3$.
\end{figure}

In the superconducting state, the spin-lattice relaxation rates $T_1^{-1}$ 
follow a thermally-activated behavior (see Fig.~\ref{fig:1_t1t}), whereas
at temperatures above $T_c$, both samples exhibit an almost linearly 
dependent $T_1^{-1}(T)$ % Temperature is already include in (T)
(see inset in Fig.~\ref{fig:korringa}).
Since both samples exhibit almost
identical relaxation rates across
the whole temperature range, this implies that carbon doping does
not significantly affect the relaxation mechanisms,
here dominated by electronics fluctuations 
related to the Pt and Zr atoms. Note that the absence of a clear 
anomaly in $T_1^{-1}(T)$ curves near the onset of SC for both compounds 
is most likely attributed to their broad superconducting transition.
In the normal state, the temperature-dependent spin-lattice relaxation 
rates $T_1^{-1}(T)$ provide useful insights into the dynamics of 
conduction electrons and their degree of correlations. 
Above $T_c$, $T_1^{-1}(T)$ deviates from a purely linear 
temperature dependence, clearly highlighted
by the $(T_1T)^{-1}$ vs $T$ plot in Fig.~\ref{fig:1_t1t}, an indication of weak- to moderate electron correlations. 
In a first approximation, the estimated value of $(T_1T)^{-1}$ at 
low temperature is about 4.72\,s$^{-1}$K$^{-1}$. 
By a comparison to diborides, such as, MgB$_2$, AlB$_2$, and ZrB$_2$ 
(also binary 
superconductors~\cite{Barbero2017}), this value indicates a relatively high electron density of states at the Fermi level 
(here dominated by the Zr and Pt $d$ bands).

Now we discuss the NMR relaxation rates in the superconducting state. 
In an $s$-wave superconductor, the opening of 
an electronic energy gap below $T_c$ implies an exponential decay 
of the NMR relaxation rate {$T_1^{-1}$}~\cite{BCS1957}: % $(T_1T)^{-1}$~\cite{BCS1957}: 
\begin{equation}
\label{eq:bcs_t1t}
%{(T_1T)}^{-1} \propto \exp\left(-\frac{\Delta_0}{k_\mathrm{B}T}\right). 
{T_1}^{-1} \propto \exp\left(-\frac{\Delta_0}{k_\mathrm{B}T}\right). 
\end{equation}
Here, $\Delta_0$ is the same as in Eq.~\eqref{eq:rhos}. 
In the superconducting state, after a slight initial enhancement,  
most likely due to the relaxation of nuclei inside the vortex
cores~\cite{Jung2001}, %$(T_1T)^{-1}$
{$T_1^{-1}$} decreases exponentially and is described very well 
by Eq.~\eqref{eq:bcs_t1t} for both 
Zr$_5$Pt$_3$ and Zr$_5$Pt$_3$C$_{0.3}$. 
It is worth noting that, at 0\,K, it converges towards zero (see 
inset in Fig.~\ref{fig:1_t1t}), thus suggesting that the electronic 
spin fluctuations represent the dominant %only
relaxation channel (reflecting 
the exponential decrease of unpaired electrons in superconducting phase). 
This result also indicates the high quality 
%
%
%==== figure =============================%
%=== end figure ==========================%
%
of the samples (i.e., a negligible concentration of 
impurities and a 100\% SC volume fraction).
\tcr{At the same time, spin fluctuations might account for the drop in
signal intensity (wipeout effect~\cite{Brom2003}) we observe below
$T_c$ (see Fig.~S6 in SM).}

The superconducting gaps $\Delta_0$ derived from 
Eq.~\eqref{eq:bcs_t1t} are listed in Table~\ref{tab:parameter}. 
The gap sizes are slightly different from the TF-$\mu$SR results, 
most likely related to the limited temperature range we can cover with 
our NMR setup. 
Although a coherence (Hebel-Slichter, HS) peak~\cite{Hebel1957} just 
below $T_c$ is a fingerprint of $s$-wave superconductivity, its 
absence not necessarily rules it out. Many factors may account for the 
suppression of the HS peak (as discussed in more detail in 
Ref.~\onlinecite{Shang2022NbReSi}), but here %Zr$_5$Pt$_3$ case 
we attribute it to the relevant contribution of $d$ orbitals at the 
Fermi level.

Finally, we consider the degree of electron correlation in the normal 
state which, in an ideal-metal case, can be deduced 
from the Korringa relation~\cite{Korringa1950}: 
\begin{equation}
\label{eq:korringa}
S = T_1TK_s^2 = \alpha S_0, \quad\text{with} \quad S_0=\frac{\gamma_e^2}{\gamma_n^2}\frac{\hbar}{4\pi k_\mathrm{B}}, 
\end{equation}
where $S$ is the Korringa constant, $\gamma_e$ is the gyromagnetic 
ratio of free electrons, and $\gamma_n$ that of the probe nucleus. 
We recall that, the total Knight shift $K$ comprises both a spin- $K_s$ 
and a (constant) orbital part $K_{\mathrm{orb}}$. 
In case of a dominant $K_s$, arising from the Fermi-contact (i.e., momentum-independent)
interaction associated with the Pauli susceptibility $\chi_\mathrm{P}$,
the Korringa relation implies an $\alpha$ close to 1.
As shown in Fig.~\ref{fig:korringa}, apart from 
a weak temperature dependence due to $T_1(T)$,  
both compounds exhibit Korringa constants %products much 
lower %higher 
than $S_{0}$ %1 
(note that, in Fig.~\ref{fig:korringa}, we report $S_ 0/S$ %$1/\alpha$ 
on the $y$-axis). 
Clearly, the assumption of an $s$-type Fermi-contact interaction, 
required for the validity of the Korringa relation, is not fulfilled 
in our case. By accounting for the Pines corrections~\cite{Pines1955} 
%p. 423, Eq. (9.18) 
to Eq.~\eqref{eq:korringa} (which include the terms $1/\rho^2(E_\mathrm{F})$ 
and $\chi^2_\mathrm{P}$, reflecting the real density of states and spin 
susceptibility), we infer that the real $\rho(E_\mathrm{F})$ due to 
$s$ electrons is less than expected.
This suggests that the relaxation has a substantial orbital character,  
proportional to the density of $d$-states at the Fermi level, but 
otherwise the correlation is low.  
%In transition metals, Korringa products $S \gg 1$ are common and reflect 
%an often substantial contribution of orbital (i.e., van Vleck) paramagnetism 
%to the Knight shift~\cite{Knight2007}. 
Therefore, we expect Zr$_5$Pt$_3$ and Zr$_5$Pt$_3$C$_{0.3}$ 
to be weakly correlated metals, despite a seemingly small 
$S$ value.

\section{Discussion}

First, we discuss why our Zr$_5$Pt$_3$C$_{x}$ samples show a different 
evolution of $T_c$ with C-content compared to a previous study, 
where the $T_c$ values first increase from 6\,K (for $x = 0$) 
to 7\,K (for $x = 0.3$).  
Then, upon further increasing the C-content, $T_c$ decreases 
continuously to 3.5\,K (for $x = 0.7$)~\cite{Renosto2019}. 
In our case, conversely, $T_c$ decreases to 4.5\,K 
already upon a 5\% C intercalation (i.e., $x = 0.05$). 
Then, upon further increasing the C-content, $T_c$ remains almost 
constant with $x$. 
Such a different evolution of $T_c$ might 
be due to a carbon content differing from the nominal one, implying 
slightly different modifications of the crystal structure.
In the previous study, upon C intercalation, the lattice was found to 
expand in the $ab$-plane, but to compress along the $c$-axis, resulting 
in a progressive decrease of the $c/a$ ratio. 
This is also the case of the Nb$_5$Ir$_{3-x}$Pt$_{x}$O family, where a smaller $c/a$ ratio leads to a higher $T_c$ value~\cite{Kitagawa2020}.     
Our Zr$_5$Pt$_3$C$_{x}$ samples, instead, show an opposite behavior, 
with the $c/a$ ratio increasing with C-content [see Fig.~\ref{fig:XRD}(b)]. 
Why our Zr$_5$Pt$_3$C$_x$ samples show a different lattice evolution 
compared to the previous work is not yet clear and requires further 
investigation.

Second, in the previous study, the temperature-dependent electronic specific heat and magnetic penetration depth (calculated from the lower critical field $H_\mathrm{c1}$) suggest a nodal superconducting gap in Zr$_5$Pt$_3$ and Zr$_5$Pt$_3$C$_{0.3}$ and possibly unconventional SC~\cite{Renosto2019}.
However, both measurements 
were limited to $\sim 2$\,K, i.e., only down to 1/3$T_c$. To reliably reveal 
the superconducting pairing, measurements at temperatures far below the 
onset of SC (i.e., at $T < 1/3T_c$) are crucial, which is also one of 
the motivations of our study.  
Therefore, the original conclusion about unconventional SC in Zr$_5$Pt$_3$C$_x$ is not solid. This is clearly demonstrated in \tcr{Fig.~S7 of SM~\cite{Supple}}, %in the Supplementary Materials~\cite{Supple}},
where we show 
the magnetic penetration depth $\lambda$ vs the reduced temperature $T/T_c$ 
for Zr$_5$Pt$_3$C$_{0.3}$, both for our sample and for that reported 
in Ref.~\onlinecite{Renosto2019}. 
As can be clearly seen, 
the $\lambda(T)$ determined from TF-$\mu$SR shows an exponential temperature dependence below $1/3T_c$. For an isotropic single-gap superconductor, the magnetic penetration depth at $T \ll T_c$
follows $\lambda(T) - \lambda_0 = \lambda_0 \sqrt{\frac{\pi\Delta_0}{2T}}e^{-\frac{\Delta_0}{T}}$~\cite{Prozorov2006},
where $\lambda_0$ and $\Delta_0$ are the same as in Eq.~\eqref{eq:rhos}. The solid line in \tcr{Fig.~S7~\cite{Supple}} is a fit to the above equation, yielding a similar superconducting gap size as that determined from superfluid density 
$\rho_\mathrm{sc}(T)$ in Fig.~\ref{fig:superfluid}. The inset in \tcr{Fig.~S7~\cite{Supple}} replots the $\lambda$ vs. $(T/T_c)^2$. As indicated by the dash-dotted lines, both data sets exhibit a $T^2$ dependence down to $T/T_c$ $\sim$ 0.3, while below it,
$\lambda(T)$ deviates significantly from the $T^2$ dependence, 
decreasing exponentially with temperature, 
consistent with a fully-gapped SC state. 
We expect also the electronic specific heat to exhibit a similar 
exponential behavior, although currently the relevant low-temperature  
data are not yet available. 
To conclude, the limited temperature range of the previous experimental 
data might lead to an incorrect conclusion about the nature of 
SC~\cite{Renosto2019}. By contrast, our TF-$\mu$SR 
measurements, performed down to 0.3\,K, i.e., well inside the 
SC state, combined with NMR results, prove beyond resonable 
doubt the conventional character of Zr$_5$Pt$_3$C$_x$ superconductivity.   

%==== Table =============================%
\begin{table}[!th]
	\centering
	\caption{ Normal- and superconducting-state properties of Zr$_5$Pt$_3$ 
	and Zr$_5$Pt$_3$C$_{0.3}$, as determined from electrical-resistivity-, 
	magnetization-, heat-capacity-, NMR-, and $\mu$SR measurements. The data 
	for Zr$_5$Pt$_3$C$_{0.5}$ were taken from Ref.~\onlinecite{Bhattacharyya2022}. 
	\label{tab:parameter}}
    \vspace{1mm}
    \renewcommand{\arraystretch}{1.12}
%	\begin{ruledtabular}
		\begin{tabular}{l@{\qquad}c@{\qquad}c@{\qquad}c@{\qquad}c}
		\toprule
			Property                                 & Unit              & Zr$_5$Pt$_3$       & Zr$_5$Pt$_3$C$_{0.3}$   & Zr$_5$Pt$_3$C$_{0.5}$    \\ \midrule %\hline
			$\rho_0$                                 & m$\mathrm{\Omega}$\,cm  & 0.095        & 0.155          & ---     \\
			$T_c^\rho$                               & K                 & 6.1                & 4.4            & 3.8     \\
			$T_c^\chi$                               & K                 & 6.3                & 4.5            & 3.8     \\
			$T_c^C$                                  & K                 & ---                & 4.1            & 3.9     \\
			$T_c^{\mu\mathrm{SR}}$                   & K                 & 6.2                & 3.9            & 3.7     \\
			$\mu_0H_\mathrm{c1}$                     & mT                & 5.5(1)             & 4.4(1)         & 5.9    \\
		    $\mu_0H_\mathrm{c1}^{\mu\mathrm{SR}}$\footnotemark[1] & mT   & 6.5(1)             & 3.2(1)         & 3.4     \\
			$\mu_0H_\mathrm{c2}$                     & T                 & 7.21(5)            & 6.97(6)        & 5.4       \\
%			$\gamma_n$                               & mJ/mol-K$^2$       & 9.6(1)            & 21.5(2)        \\
%			$\Theta_\mathrm{D}$                      & K                  & 390(3)            & 440(5)         \\
%			$N(\epsilon_\mathrm{F})^C$               & states/eV-f.u.     & 4.1(1)            & 9.1(1)         \\
%			$N(\epsilon_\mathrm{F})^\mathrm{DFT}$    & states/eV-f.u.     & 2.35              & 5.7            \\ %
%			%	$E_\mathrm{SOC}^\mathrm{DFT}$            & meV                & xxx           & xxx            \\[2mm]
			$\Delta_0^{\mu\mathrm{SR}}$\footnotemark[2]  & meV           & 1.20(2)            & 0.60(2)        & 0.59 \\
			$\Delta_0^{\mu\mathrm{SR}}$\footnotemark[3]  & meV           & 1.17(2)            & 0.47(2)        & ---  \\
%        		$\Delta_0^{\mathrm{NMR}}$                & meV               & 0.81(8)            & 0.91(8)        & ---  \\
			$\Delta_0^{\mathrm{NMR}}$                & meV               & {0.98(8)}      & {1.08(8)}        & ---  \\
%			$\Delta_0$($\mu\mathrm{SR}$)$^\mathrm{clean}$             & meV               & 0.77(2)        & 0.50(2)           \\  
%			$\Delta_0$($\mu\mathrm{SR}$)$^\mathrm{dirty}$             & meV               & 0.66(2)        & 0.44(1)           \\  
%			$w$                                      & --                  & 0.7          & 0.27           \\
%			$\Delta_0^f(C)$                          & meV                & 0.69(2)       & 0.32(1)           \\  
%			$\Delta_0^s(C)$                          & meV                & 0.79(2)       & 0.50(1)           \\  
%			$\Delta_0^f(\mu\mathrm{SR})$             & meV                & 0.72(1)       & 0.35(1)           \\  
%			$\Delta_0^s(\mu\mathrm{SR})$             & meV                & 0.87(2)       & 0.57(2)           \\  
			$\xi(0)$                                 & nm                 & 6.10(3)          & 6.25(2)       & 7.81     \\
            $\kappa$                                 & ---                & 61               & 67            & 60       \\ 
			$\lambda_0$                              & nm                 & 333(3)           & 493(2)        & 469      \\ %[2mm]
			$\lambda_\mathrm{GL}(0)$                 & nm                 & 366(4)           & 415(6)        & ---      \\
			$\Lambda_\mathrm{ZF}$(0.3\,K)            & $\mu$s$^{-1}$      & ---              & 0.018(2)     & ---      \\
		    $\sigma_\mathrm{ZF}$(0.3\,K)             & $\mu$s$^{-1}$      & ---              & 0.046(8)      & ---       \\ 
		   	$\Lambda_\mathrm{ZF}$(10\,K)             & $\mu$s$^{-1}$      & ---              & 0.015(4)     & ---      \\
		    $\sigma_\mathrm{ZF}$(10\,K)              & $\mu$s$^{-1}$      & ---              & 0.048(8)     & ---       \\ \bottomrule
%			$\lambda_\mathrm{L}$                      & nm                 & 90(5)         & 229(2)            \\
%			$l_\mathrm{e}$                           & nm                 & 2.2(1)        & 22(1)             \\
%			$\xi_0$                                  & nm                 & 17(1)         & 6.4(1)             \\
%			$\xi_0$/$l_\mathrm{e}$                   & --                 & 7.7           & 0.3                \\ %
%			$m^{\star}$                              & $m_e$              & 7.0(2)        & 10.4(1)            \\
%			$n_\mathrm{s}$                           & 10$^{28}$\,m$^{-3}$ & 2.4(3)       & 0.56(1)            \\ 
%			$v_\mathrm{F}$                           & 10$^5$\,ms$^{-1}$   & 1.5(1)       & 0.61(1)            \\
%			$T_\mathrm{F}$                           & 10$^4$\,K           & 1.0(1)       & 0.25(1)            \\
			%	$\lambda_\mathrm{GL}$                  & nm              & 232(4)     & 107(1)  \\
		\end{tabular}	
%	\end{ruledtabular}
     \footnotetext[1]{Calculated by $\mu_{0}H_\mathrm{c1} = (\Phi_0 /4 \pi \lambda^2)[$ln$(\kappa) + 0.497]$.}
     \footnotetext[2]{Derived from clean-limit model.}
      \footnotetext[3]{Derived from dirty-limit model.}
     %\footnotetext[2]{ZF-$\mu$SR relaxation at 0.3\,K.}
\end{table}
%=== end table ==========================%
%

Third, the $\mu$SR and NMR data of Zr$_5$Pt$_3$ and Zr$_5$Pt$_3$C$_{0.3}$ presented here, together with the previous study on Zr$_5$Pt$_3$C$_{0.5}$, suggest that the conventional SC of Zr$_5$Pt$_3$C$_{x}$ is independent of carbon 
content. Unlike the Nb$_5$Ir$_{3-x}$Pt$_x$O family, where a crossover from 
multiple- to single-gap SC is observed upon Pt 
doping~\cite{Kitagawa2020,Xu2020}, here, the single-gap SC in 
Zr$_5$Pt$_3$C$_{x}$ is robust against the intercalation of interstitial 
carbon atoms. 
This is most likely attributed to their similar electronic band structures. According to DFT calculations, up to six bands cross the Fermi level~\cite{Bhattacharyya2022}. The density of states (DOS) is dominated by 
Zr $d$-orbitals (up to 68\% of the total DOS, the rest being due to Pt), 
while the contribution of C $p$-orbitals is negligible. Moreover, the 
change in DOS is only about 10\%, even when increasing the C-content up to $x = 1$.   

\section{Conclusion}
To summarize, we investigated the normal- and superconducting properties
of Zr$_5$Pt$_3$C$_x$ ($x = 0$--0.3) family of compounds by means of 
electrical resistivity-, magnetization\mbox{-,} heat capacity-, $\mu$SR, and NMR
measurements. The Zr$_5$Pt$_3$C$_x$ exhibit bulk SC with $T_c$ between 
4.5\,K and 6.3 K. The electrical-resistivity measurements under applied 
magnetic field reveal a zero-temperature upper critical field of 
$\sim$7\,T. 
The temperature dependence of the superfluid density reveals a 
\emph{nodeless} SC in Zr$_5$Pt$_3$C$_x$, well described by an isotropic 
$s$-wave model. The conventional nature of Zr$_5$Pt$_3$C$_x$ 
superconductivity is also 
supported by the exponential temperature-dependent NMR relation rate 
and the change 
of Knight shift below $T_c$. 
Finally, the lack of spontaneous magnetic
fields below $T_c$ indicates that the time-reversal 
symmetry is preserved in the Zr$_5$Pt$_3$C$_x$ superconductors.

\vspace{3pt}%
\begin{acknowledgments}
	This work was supported from the Natural Science Foundation of 
	Shanghai (Grant Nos.\ 21ZR1420500 and 21JC1402300) and the Schweizerische 
	Nationalfonds zur F\"{o}rderung der Wis\-sen\-schaft\-lichen For\-schung 
	(SNF) (Grant Nos.\ 200021\_188706 and 206021\_139082). We acknowledge the allocation of beam time at the Swiss muon source, 
	and thank the scientists of Dolly $\mu$SR spectrometer for their support.
\end{acknowledgments}

\bibliography{Zr5Pt3_bib}
%\end{footnotesize}

\end{document}